%% file: paper.tex
\documentclass[nofootinbib,twocolumn,prd]{revtex4}
\usepackage{amsmath}
\usepackage{graphicx}
\usepackage{color}
\usepackage{ifpdf}
\usepackage{longtable}

\newcommand\editremark[1]{ {\color{red} #1}}

\newcommand\hidetosubmit[1]{}
\newcommand\optional[1]{}
\newcommand\ForInternalReference[1]{}

\newcommand\Y[1]{Y^{(#1)}}

\newcommand\avL{\left< {\cal L}_{(a} {\cal L}_{b)} \right>}
\newcommand\WeylScalar{{\psi_4}}
\newcommand\WeylScalarFourier{{\tilde{\psi}_4}}

\newcommand\qmstate[1]{\left|#1\right>}
\newcommand\qmstateKet[1]{\left<#1\right|}
\newcommand\qmstateproduct[2]{\left<#1|#2\right>}
\newcommand\qmoperatorelement[3]{\left<#1\left|#2\right|#3\right>}

\newcommand{\ROS}[1]{#1}

\def\bbh#1{binary black hole#1 (BBH#1)\gdef\bbh{BBH}}
\def\bh#1{black hole#1 (BH#1)\gdef\bh{BH}}

\newcommand\chiVector{\vec{{a}}}

\begin{document}
\title{Precession during merger 1: Strong polarization changes are observationally accessible features of strong-field gravity
  during binary black hole merger}
\author{R.\ O'Shaughnessy}
\affiliation{Center for Gravitation and Cosmology, University of Wisconsin-Milwaukee,
Milwaukee, WI 53211, USA}
\email{oshaughn@gravity.phys.uwm.edu}
 \author{ L. London}
 \author{ J. Healy}
 \author{D. Shoemaker}
 \affiliation{Center for Relativistic Astrophysics,
 Georgia Tech, Atlanta, GA 30332, USA}

\begin{abstract}
The short gravitational wave signal from the merger of compact binaries encodes a surprising amount of information about
the strong-field dynamics of merger into frequencies accessible to ground-based interferometers.
In this paper we  describe a previously-unknown ``precession'' of the peak emission direction with time, both before and \emph{after}
the merger, about the total angular momentum direction.   
We demonstrate that the gravitational wave polarization encodes the orientation of this direction to the line of sight.
We argue that the effects of polarization can be estimated \emph{nonparametrically}, directly from the gravitational wave signal as seen along one line
of sight, as a slowly-varying feature on top of a rapidly-varying  carrier. 
After merger, our results can be interpreted as a coherent excitation of quasinormal modes of different angular orders, a superposition which naturally
``precesses'' and modulates the line-of-sight amplitude.        
Recent analytic calculations have arrived at a similar geometric interpretation.
We suspect the line-of-sight polarization content will be a convenient observable with which to define  new
high-precision tests of general relativity using gravitational  waves. 
 Additionally, as the nonlinear merger process seeds the initial coherent perturbation, we speculate that the
\emph{amplitude} of this effect provides  a new probe of the strong-field dynamics during merger.  
To demonstrate that the ubiquity of the effects we describe, we summarize  the post-merger evolution of 104 generic precessing binary mergers.
Finally, we provide  estimates for the detectable impacts of precession on the waveforms from
high-mass sources.     These expressions may  identify new precessing binary parameters whose waveforms are dissimilar from the existing sample.
\end{abstract}

\maketitle

\ForInternalReference{
\editremark{Before submission}
- collect comments
- confirm numbers in table
- eliminate superfluous text.  
- Rotation bugfix: Update text to refer to J when necessary.  Make sure figure captions are rotated by 180 degrees

}

\ForInternalReference{

NAME CHECKED PUBLICATIONS

- Jim/Pablo, kicks

- Larne, banksim with precession.  Model the mismatch

- Jim, describe simulations. 

- Corotating waveforms

- Selection bias

}

\hidetosubmit{\editremark{Technical improvements}: Schmidt's decomposition; superoperators; other tricks from qm
  information
}

\optional{

a) significant change in band of direction 

b) AND we're lucky enough to be oriented not head on, so we see it.  (not insignificant chance, easier with big changes)

...then, we can measure more or less a deflection angle and shape

}

\section{Introduction}

\label{sec:Intro}

Coalescing comparable-mass black hole binaries are among the most likely and useful sources of gravitational waves for existing and
planned gravitational wave detectors like LIGO \cite{gw-detectors-LIGO-original-preferred},  Virgo,
\cite{gw-detectors-Virgo-original-preferred}, the Einstein telescope \cite{2010CQGra..27s4002P}, and proposed space-based detectors.  
For sources in a suitable mass range, the signal these detectors receive contains significant
features from the late-stage, strong-field  dynamics of the black hole merger.
Only full numerical simulations of Einstein's equations can provide first-principles models for this epoch, including
all dynamics and emission  \cite{Lehner:2001wq,2010RvMP...82.3069C}.  %
Given the large computational cost per simulation, relatively few  well-determined models have been
produced.\footnote{Most simulations have thoroughly explored  nonspinning and spin-aligned binaries \cite{2010RvMP...82.3069C}.  Some simulations with more generic spin have been
performed (e.g., to study gravitational wave recoil kicks)
\cite{Herrmann:2007ex,Herrmann:2007ac,Healy:2008js}.}

For sufficiently high mass mergers,  ground-based gravitational wave detectors are sensitive only to the  short portion of
the waveform  provided by numerical simulations.    
Historical comparisons of the leading-order emission from comparable-mass nonprecessing binaries has suggested these
waveforms look similar
 \cite{gr-merger-fitting-Alessandra-2006,2010RvMP...82.3069C}.  
 Existing upper limits on gravitational waves from
$>50 M_\odot$ merging binaries  imply astrophysically plausible signal amplitudes for the first few detections must be
 low, at best near the detection limit of the advanced LIGO and Virgo detectors
\cite{LIGO-Inspiral-S6-Highmass}.   
 Short, similar, low amplitude signals can
carry a limited amount of information.   
Nonetheless, strong-field dynamics produces significant evolution in the emission beampattern and polarization content
during merger.  If in band, these features should be accessible to gravitational wave detectors.

In this paper we provide the first phenomenological description of the ``kinematics of merger''.  Rather than discuss
properties of horizons or of the global spacetime \cite{2011PhRvL.106o1101O,2011PhRvD..84l4014N}, we emphasize
phenomenology of the asymptotic radiation.  In particular, we identify precession-induced changes in the polarization
content as an easily-accessible phenomenological measurable with which to distinguish between simulations.    While we
adopt a simple model to characterize this diagnostic, our results can be  translated to fully-developed parameter
estimation and model selection strategies, which fit data to a model using physically-motivated
\cite{1998PhRvD..58h2001C,2008PhRvD..78b2001V,2007CQGra..24..607R,2007PhRvD..75f2004R,2008ApJ...688L..61V,2008CQGra..25r4011V}
and phenomenological  \cite{2011PhRvD..84f2003C,gr-extensions-tests-Europeans2011,2011PhRvD..83h2002D}  signal parameters.
For clarity we use three  examples to illustrate the precession we have in mind and the imprint it makes on
emitted signals.    To demonstrate that the features we describe occur frequently and to connect those features to
  quasinormal mode frequencies, we also describe the post-merger  dynamics of a much larger array of simulations.  
In Section \ref{sec:Simulations} we describe the gravitational wave signals, preferred orientations, and description of polarization we  use in
our work.   We argue that our practical approach to gravitational wave polarization, decomposing the line of sight signal
into left- and right-handed components, is both observationally accessible and intuitive.   
Our study uses detailed numerical relativity simulations of the late inspiral and merger of two  black
holes, in general both with spin.     By example, we demonstrate that ``precession'' dynamics, suitably defined,
continues after merger.   Again using selected examples, we illustrate that \emph{each line of sight} encodes the
relative orientation of the (time dependent, precessing) preferred orientation.  
By comparing these  to a simple expression, we argue this behavior occurs in general, for all lines of sight and all
precessing binaries.  
In Section \ref{sec:Simulations2} we demonstrate that all these features change significantly, depending on the spin
magnitudes and orientations adopted.   Based on these investigations, we anticipate the polarization content can trace
features of the strong-field merger event.
Then, in Section \ref{sec:PolarizationImprint}, we demonstrate that this polarization information is experimentally accessible: similar
waveforms differing primarily through their polarization content (i.e., time history of left- versus right-handed) 
can be observationally distinguished. 
We demonstrate that observations can weakly constrain orientation evolution \emph{nonparametrically} , independent of
other information naturally encoded in the gravitational wave signal.

\subsection{Context }

Though the gravitational-wave signal from merging binaries nominally encodes all information about the spacetime, 
previous investigations of the $(2,2)$ mode suggest  universality of the  merger signal  \cite{gr-merger-fitting-Alessandra-2006,2010RvMP...82.3069C}, giving
the informal impression that  previously hypothesized  complicated nonlinearities in general relativity do not
significantly complicate the  short signal from the merger epoch.  In other words, simulations seemed to suggest that the merger signal provides little unambiguous information about
strong-field general relativity.
However, the unique contribution of merger to the waveforms is invariably mixed with the preceding (inspiral) and
subsequent (ringdown) signal, making nonlinear behavior difficult to identify.  In principle  the mass loss, angular
momentum change, and recoil kick can all be tabulated, fit, and constrain the nonlinear response at merger.
Nevertheless, in practice the contribution from merger cannot be cleanly distinguished from the contributions from
inspiral and ringdown.  While symmetry demands a general form  \cite{gr-nr-io-fitting-Boyle2007}, it does not strongly forbid inspiral or merger
epochs from contributing to these integrated quantities.
The contribution from merger can be partially disentangled in the time domain, for example by plotting fluxes like $dE/dt$, or possibly
the time-angular domain.
Unfortunately, investigations looking for strong mode-mode couplings indicative of nonlinearity have recovered the
familiar perturbative results \cite{2003PhRvD..68h4014Z,2007PhRvD..76h4007N,2010PhRvD..82j4028P}.  
Previous investigations suggest that the asymptotic radiation encodes preferred directions with \emph{different} symmetry
properties, not tied to the orientations associated with standard conserved constants.  These orientations are well
resolved through merger and seem to evolve nontrivially both before and after the merger
\cite{gwastro-mergers-nr-Alignment-ROS-Methods,gwastro-mergers-nr-Alignment-ROS-IsJEnough}.
This preferred orientation may therefore provide a new signature of strong-field dynamics.

After this study was completed, we became aware of an analytic study providing a simple, geometric interpretation of
(short-wavelength) quasinormal modes' frequencies as precession
\cite{gwastro-YangZimmermanEtc-QuasinormalSpectrumAndPrecession2012}.  
Their calculation extended previous  estimates derived in the slow-rotation limit to arbitrary spin.    
We find a similar result at the lowest multipolar order, even in the strong field: all modes precess at
a nearly constant rate during and after merger.

\section{Simulations and diagnostics}
\label{sec:Simulations}

\subsection{Simulations I: Overview}
The simulations examined in this paper will be described at greater length in a companion publication.   
Initial binary properties and
simulation details including initial separation, spin configuration, and 
finest resolution are detailed in Table \ref{tab:Simulations:VPrecession}.
 Physically, roughly half of these binary
configurations correspond to an unequal-mass generalization of the ``S series''  \cite{Herrmann:2007ex}, where one spin is in the orbital
plane $\chiVector_2 = a \hat{x}$ and the other spin rotates through the $(x,z)$ plane with comparable dimensionless spin
$a=|{\vec S}_1/M^2_1|=|{\vec S}_2/M^2_2|$.   
These simulations are denoted by the prefixes S and Sq in Table \ref{tab:Simulations:VPrecession}.   
\ROS{The remaining simulations (T and Tq) correspond physically to equal and unequal-mass binaries where the less massive object has its spin
\emph{parallel} to the initial orbital angular momentum ($\chiVector_{1\text{ or } 2} = a \hat{z}$), while the more massive
object's spin rotates through the $(x,z)$ plane with comparable dimensionless spin.\footnote{For the equal-mass T
  series, object 2 rather than object 1 had its spin tilted.}
}
Initial data were evolved with  \texttt{Maya}, which was used in previous \bbh{} studies \cite{Herrmann:2007ex,Herrmann:2007ac,Hinder:2007qu,Healy:2008js,Hinder:2008kv,Healy:2009zm,Healy:2009ir,Bode:2009mt}.
The grid structure for each run consisted of  10 levels of refinement 
provided by \texttt{CARPET} \cite{Schnetter-etal-03b}, a 
mesh refinement package for \texttt{CACTUS} \cite{cactus-web}. 
Sixth-order spatial finite differencing was used with the BSSN equations 
implemented with Kranc \cite{Husa:2004ip}. 
In the text, we present results for the fiducial simulation  at $r=90 M$ unless otherwise indicated.   The choice of
extraction radius has minimal impact on the results shown here.  
Finally, our code projects $\psi_4(r,t)$ onto spin-weighted spherical harmonics;  we adopt a phase convention such that
$\Y{s}_{lm}(\theta,\phi)e^{is \gamma} = (-1)^s \sqrt{(2l+1)/4\pi} d^l_{m,-s}(\theta) e^{im \phi}e^{i s \gamma}$ \cite{2007JCoPh.226.2359W}.  

\ROS{ While our principal interest remains observationally accessible radiation at infinity, for illustration  we also
  estimate and plot the black hole spins and orbital angular momentum as a function of time.   Lacking any
  gauge-invariant definition, not having computed $\vec{S}_{1,2}$ for all time for all simulations, and not intending to
  perform quantitative comparisons between our results and post-Newtonian expressions for $\hat{L}$ that differ from Newtonian expressions only at $v^3$ order
  and above, we simply adopt the \emph{coordinate} orbital angular momentum, calculated from the two puncture
  locations $r_1, r_2$:
$
\vec{L}\equiv \sum_k M_k \vec{r_k}\times   \partial_t \vec{r}_k
$.
As in previous Maya simulations \cite{Herrmann:2007ex}, we extract the black hole spins from (a small neighborhood surrounding the) apparent
horizon, using a surface integral as if the horizon was isolated  \cite{2003PhRvD..67b4018D,2004LRR.....7...10A,2006PhRvD..74b4028S}.
}

\begin{figure}
\includegraphics[width=\columnwidth]{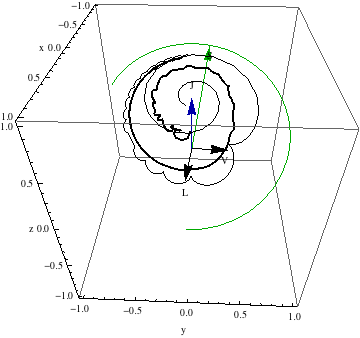}
\caption{\label{fig:OrientationsVersusTime:Fiducial}\textbf{Evolution of angular momenta and $\hat{V}$ with time}:
 For
  one simulation (Tq(4,0.6,90), with $M_1/M_2=4$, $a_1=0.6 \hat{x}$, $a_2=0.6 \hat{z}$ starting at $d=10M$), a plot of
  the preferred direction $\hat{V}$ (thick
  black; see Eq. (\ref{eq:def:avL})) superimposed on the coordinate orbital (black) and  spin (green) angular momenta directions
  $\hat{L},\hat{S}_1$ as a function of time.   The paths indicate how all three evolve with time; the arrows are
  evaluated at a specific instant %
shortly before merger.  
A blue arrow also indicates the final ($\simeq$ initial) total angular
  momentum $J$.     At very early times, the directions $\hat{V}$ and $\hat{L}$ nearly agree \cite{gwastro-mergers-nr-Alignment-ROS-PN}; at merger, however,
  $\hat{L}$ converges to $\hat{J}$ while $\hat{V}$ continues to precess.  
   Compare to figures in \citet{gwastro-mergers-nr-Alignment-ROS-IsJEnough}.
\ROS{  This simulation is longer than but physically similar to the fiducial simulation adopted in all subsequent
  figures (Sq(4,0.6,90,9)).}
}
\end{figure}

To  illustrate features of these simulations, we will emphasize one fiducial binary, denoted Sq(4,0.6,90,9)
with  $ M_1/M_2=4$, $\chiVector_1=0.6 \hat{x} = -\chiVector_2$,
placed in orbit  in the $x,y$ plane wth a coordinate separation $d=9M$; see Table \ref{tab:Simulations:VPrecession} for details.\footnote{Our fiducial initial conditions are
  comparable to those adopted in a companion paper on orientation-dependent emission \cite{gwastro-mergers-nr-Alignment-ROS-IsJEnough}.}
For a very similar simulation [T2(4,0.6,90)], Figure \ref{fig:OrientationsVersusTime:Fiducial} shows the  inspiral-merger trajectories of  the orbital angular momentum and spin directions, $\hat{L}(t)$ and $\hat{S}_1(t)$, where $\hat{S}_1(t)$ corresponds to the larger black hole.    
Initially, the binary's orbital angular momentum is larger but comparable to $\vec{S}_1=\vec{a}_1M_1^2$; both are much
larger than the companion's spin.
 Early in the inspiral, $\hat{L}(t)$ and $\hat{S}_1(t)$ precess around the total angular momentum direction \cite{ACST}.   During this process, the three  angular momenta ($\vec{J},\vec{L}$ and $\vec{S}_1$) are nearly coplanar, with the angle
between $\vec{L}$ and $\vec{S}_1$ roughly constant,  as expected from the post-Newtonian limit.   
By contrast, as the orbit transitions from inspiral to plunge, the (coordinate) orbital angular momentum \emph{converges} towards
the total angular momentum direction.  The spin direction, by contrast, changes far less dramatically during the plunge
and merger.

\subsection{Preferred orientations}

Conserved quantities of the spacetime and radiation escaping to infinity encode several preferred orientations.   One
preferred, manifestly physical, but nearly-time-independent  orientation is $\hat{J}$, the nearly conserved orientation of the total angular momentum.  
Another orientation extremizes the quadrupolar radiation, such that  the $(2,2)$ mode of the radiative Weyl scalar ($\WeylScalar$) is largest
\cite{gwastro-mergers-nr-ComovingFrameExpansionSchmidt2010,gwastro-mergers-nr-Alignment-BoyleHarald-2011}; we denote this orientation as    $\hat{Q}(t)$.
Finally, a third orientation $\hat{V}(t)$ is the principal axis of \cite{gwastro-mergers-nr-Alignment-ROS-Methods,gwastro-mergers-nr-Alignment-ROS-IsJEnough,gwastro-mergers-nr-Alignment-ROS-PN}
\begin{eqnarray}
\label{eq:def:avL}
\avL_t &\equiv& 
 \frac{\int d\Omega \WeylScalar^*(t) {\cal L}_{(a}{\cal L}_{b)} \WeylScalar(t)
  }{
   \int d\Omega |\WeylScalar|^2
}
 \\
 &=& \frac{\sum_{lmm'} \WeylScalar_{lm'}^*  \WeylScalar_{lm}\qmoperatorelement{lm'}{{\cal L}_{(a}{\cal L}_{b)}}{lm} }{\int d\Omega |\WeylScalar|^2}  \nonumber
\end{eqnarray}
where ${\cal L}$ are rotation group generators; where $\WeylScalar$ is the Weyl scalar; and
where in the second line we expand $\WeylScalar= \sum_{lm} \WeylScalar_{lm}(t)\Y{-2}_{l,m}(\theta,\phi)$ and perform the
angular integral; see \citet{gwastro-mergers-nr-Alignment-ROS-PN} for explicit formulae.

These three  quantities do not agree.  For example, early in the inspiral $\hat{J}$ is manifestly distinct from the two
precessing, dynamic orientations  $\hat{V},\hat{Q}$, which both nearly correspond to $\hat{L}$
\cite{gwastro-mergers-nr-ComovingFrameExpansionSchmidt2010,gwastro-mergers-nr-Alignment-ROS-PN}. 
The two dynamic orientations also generally differ, albeit less significantly.\footnote{While the differences between
  the two approaches to preferred orientations are often small in \emph{absolute} scale, these small angular differences
  imply significant disagreement about the relative magnitude of strongly suppressed modes, particularly in the merger phase.}
As a concrete example, if the Weyl scalar consists only of the $l=2$ subspace with the time-independent value
\begin{eqnarray}
\WeylScalar=\Y{-2}_{2,2}+\Y{-2}_{2-,2} + (\Y{-2}_{2,1}+\Y{-2}_{2,-1})/4 +  \Y{-2}_{2,1}/3
\end{eqnarray}
then on the one hand our preferred direction has $\hat{V}\simeq \hat{n}(0.045, 0)$ but
the direction that maximizes the $(2,2)$ mode lies along $\hat{n}(0.66,0)$, where $\hat{n}(\theta,\phi)$ is a cartesian
unit vector with polar coordinates $\theta,\phi$. 
For simplicity, in this work we only present results for a single  dynamic orientation $\hat{V}$. 

Figure  \ref{fig:OrientationsVersusTime:Fiducial}  %
 shows the trajectories of $\hat{L},\hat{V}$ and $\hat{J}$. %
 While $\hat{L}$ and $\hat{V}$ nearly agree prior to merger,  they differ significantly at and after merger.   
 In the time domain, the direction $\hat{V}$ roughly corresponds to the
direction of peak $|\WeylScalar|$.   More critically, when averaged over mass and frequency, this direction corresponds
very closely to the direction of largest signal amplitude
\begin{eqnarray}
\label{eq:def:rho}
\rho^2(\hat{n})&\equiv& 2\int_{-\infty}^{\infty} \frac{df}{S_h} \frac{|\psi_4(f,\hat{n})|^2}{(2\pi f)^4}
\end{eqnarray}
where we adopt the isotropic two-detector network definition of $\rho^2(\hat{n})$ used in
\citet{gwastro-mergers-nr-Alignment-ROS-IsJEnough}. 
Moving forward in time, both $\hat{V}$ and $\hat{L}$  precess around the total angular momentum $\hat{J}$ \cite{ACST}.
When the two black holes merge, the orbital angular momentum %
has converged to the total angular momentum
($\hat{L}\rightarrow \hat{J}$).  This direction no longer evolves past merger.
By contrast, the observationally-relevant direction $\hat{V}$ continues to evolve through and beyond merger.

 As will be demonstrated using the fiducial simulation, different preferred directions identify different physics.  In particular, $\hat{V}$ is  directly connected to observations, and indicates interesting new dynamics that could be
imprinted on the gravitational wave signal.

Here, we identified this imprint at each time using \emph{all gravitational wave emission directions}.    Experiments have
access to only one line of sight.  
Below, we describe a concrete method to identify the imprint of $\hat{V}$ using only a single emission direction, via the
polarization evolution versus time. 

\subsection{Line of sight diagnostics: amplitude, phase, polarization }

Real gravitational wave detectors only have access to  a single line of sight.   Despite this limitation, they still
have access to the rich and distinctive  modulations that the orbit and precession imparts on the line-of-sight 
time-evolving gravitational wave signal, $h\equiv h_++i h_\times$.  
To interpret these modulations, we change basis from the standard linear polarizations ($h_+,h_\times$) to  two circular
polarizations ($h_R,h_L$ with $h=h_R+h_L$).     Strictly, we split the Weyl scalar
$\WeylScalar$ rather than $h(t)$, to minimize errors from poorly-constrained early-time and late-time
effects.\footnote{We anticipate the gravitational wave content of the $(2,2)$ mode to be almost exclusively
  right-handed.  However, the linear transformation from $h\rightarrow h_{R,L}$ is highly nonlocal, depending in principle on
  arbitrarily early and late times.  The gravitational wave strain at the start (inspiral) or end (nonlinear memory) can
  be significant.  By contrast, the Weyl scalar's diminished early-time amplitude and lack of memory terms leads to a
  well-behaved projection.}
We will see below that precession distinctively modulates the amplitude and phase of both circular polarizations.

The left and right-handed circularly polarized parts of $\WeylScalar(t)$ are naturally isolated in the frequency domain
\cite{gwastro-mergers-nr-Alignment-ROS-IsJEnough}: positive frequency components are right-handed, and negative frequency components are left-handed.
In the time domain, this procedure corresponds to  convolution of $\WeylScalar$ with a specific (acausal) kernel:
\begin{subequations}
\label{eq:Chiral}
\begin{eqnarray}
\WeylScalar_R &=& %
\int_0^{\infty} df \WeylScalarFourier(f) e^{-2\pi i f t} \\
 &=& \int_{-\infty}^{\infty} d\tau K_R(t-\tau) \WeylScalar(\tau) \\
K_R(\tau) &=& \lim_{\epsilon \rightarrow 0^+} \frac{-i }{\tau - i \epsilon} \frac{1}{2\pi} 
\\
\WeylScalar_L &=& %
\int_{-\infty}^{0} df \WeylScalarFourier(f) e^{-2\pi i f t} \\
K_L(\tau) &=& \lim_{\epsilon \rightarrow 0^+} \frac{+i }{\tau + i \epsilon} \frac{1}{2\pi}  %
\end{eqnarray}
\end{subequations}
\ForInternalReference{This relation is extensively   used in QM and linear response theory.}
where   $K_{R,L}(\tau)$ are suitable  inverse fourier transforms of the unit step function.
For each polarization, we can define an amplitude $|h_R|$ and phase $\text{arg}(h_R)$.  In general, the amplitude evolves
on a precession (or, if none, radiation reaction) timescale;  the phase increases secularly, on a radiation reaction
timescale, with precession-induced modulations on shorter timescales \cite{ACST}.
This procedure for separating left- and right-handed signals can be perfomed with real gravitational wave data.  
For each line of sight, experiments with comparable sensitivity to two linear polarizations can reorganize their data analysis procedure to be
sensitive to only these  left- or right-handed polarizations.     
Additionally, this projection process has physically expected properties, applied to nonprecessing binaries.   The
angular modes $\WeylScalar_{lm}$ describe emission preferentially above ($m>0$) or below ($m<0$) the orbital plane,
where $m$ is the mode order (i.e., each term in $\WeylScalar=\sum_{lm}\WeylScalar_{lm} \Y{-2}_{l,m}(\theta,\phi)$'s mode decomposition
is proportional to some $ \WeylScalar_{lm}\exp{im\phi}$).
One can empirically verify that the angular modes from nonprecessing binaries are nearly chiral:
\begin{eqnarray}
\WeylScalarFourier{}_{lm}(f)\simeq 0 \quad \text{for} \quad  m f< 0
\end{eqnarray}
In the stationary-phase limit, this relationship corresponds to a desirable and usually satisfied requirement on the angular frequency
versus time:\footnote{Due to projection effects of spheroidal harmonics onto spherical harmonics, this property need not
  hold for all higher-order modes.  We will not address angular mode mixing in this paper.}  $\partial_t \text{arg}\WeylScalar_{lm}$ is monotonically increasing for $m>0$ and decreasing for
$m<0$. 
For nonprecessing binaries, to a good approximation our polarization projection corresponds to eliminating modes of
either $m<0$ (projecting to $R$) or $m>0$ (projecting to $L$).

To decouple polarization content from secular trends in amplitude and phase seen in nonprecessing binaries, we once again change
variables from $\WeylScalar_{R,L}$ (2 complex or 4 real quantities) to a real typical amplitude $A$, a real typical
phase $\Phi$,
and a complex polarization amplitude $z_\psi$ (one complex and two real quantities):
\begin{eqnarray}
 z_\psi & \equiv &\frac{\WeylScalar_L^*}{\WeylScalar_R}  \\
A^2 &\equiv& |\WeylScalar_L|^2 + |\WeylScalar_R|^2  \\
e^{i \Phi} &\equiv& \left(\frac{\WeylScalar_L^*}{\WeylScalar_L}\frac{\WeylScalar_R}{\WeylScalar_R^*} \right)^{1/2}
\end{eqnarray}

By construction, the polarization amplitude $z_\psi$ is nearly constant in time for nonprecessing binaries whose gravitational wave
emission is dominated by equal in magnitude and conjugate in phase $(l,m)=(2,2)$ and  $(2,-2)$ modes.    In this limit, the Weyl scalar
asymptotically takes the form 
\begin{eqnarray}
\label{eq:Psi4ToyModel:22Only}
r \WeylScalar(\hat{n},t) =  \frac{A(t)}{\sqrt{2}} [e^{i  \Phi(t)} \Y{-2}_{2,2}(\hat{n}) + e^{-i  \Phi(t)} \Y{-2}_{2,-2}(\hat{n})]
\end{eqnarray}
To recover the correct limits along the $\hat{z}$ axis (i.e., the $\pm \hat{z}$ direction), the basis coefficients of the two spin-weighted harmonics
$\Y{-2}_{2,\pm2}$  are necessarily right and left circularly polarized.  As a result, for a signal dominated by equal and
conjugate $(2,2)$ and $(2,-2)$ modes in a frame aligned with $\hat{L}$, the polarization amplitude is a time-independent and purely geometrical
expression:
\begin{align}
\label{eq:z:ExampleNonprecessingExtreme}
z_\psi &\simeq& \Y{-2}_{2-2}{}^*(\hat{n}|\hat{L}) /\Y{-2}_{2,2}(\hat{n}|\hat{L}) = e^{i 4\Psi_L}  \frac{(1-\cos \theta)^2}{(1+\cos \theta)^2}
\end{align}
where $\Y{-2}_{l,m}(\hat{n}|V)$ are spin-weighted harmonics in a frame aligned with $\hat{V}$ and where  $\theta =\cos^{-1} \hat{L}\cdot \hat{n}$ and $\Psi_L$ characterize the orientation of the orbital angular momentum
along the line of sight and in the plane of the sky, respectively; see, e.g. \cite{gw-astro-SpinAlignedLundgren-FragmentA-Theory}.

Reversing the sense of the approximation above, assuming  instantaneous and symmetric $(2,\pm 2)$ emission along a proposed
slowly-varying orientation $\hat{O}(t)$ , we  estimate the
polarization content $\tilde{z}_{O}(t)$ along a line of sight $\hat{n}$ by
\begin{eqnarray}
\label{eq:z:ExampleReconstruct:SlowPrecessing}
\tilde{z}_O &=& \frac{[\hat{O}\cdot (\hat{x}_n+i \hat{y}_n)]^4}{ (1- (\hat{O}\cdot \hat{n})^2)^2}
  \frac{  (1- \hat{O}\cdot \hat{n})^2}{(1+\hat{O}\cdot \hat{n})^2}  \nonumber \\
 &=& \frac{[\hat{O}\cdot (\hat{x}_n+i \hat{y}_n)]^4}{ (1+ \hat{O}\cdot \hat{n})^4}
\end{eqnarray}
where the two vectors $\hat{x}_n,\hat{y}_n$ define a frame of reference in the plane of the sky, perpendicular to $\hat{n}$.
As this simple approximation suggests, for a precessing binary the polarization amplitude $\tilde{z}_O$ should
fluctuate, reflecting the relative importance and phasing of  left- and right-handed emission in the instantaneous signal.

Closer to the orbital plane, the balance of polarizations is nearly equal ($|z_\psi|\simeq 1$).  This naturally finely-tuned
region involves near-perfect cancellation of (some of) the leading-order emission,\footnote{Near the oribital plane of
  a nonprecessing binary, gravitational wave emission is nearly linearly polarized.} allowing higher-order angular
dependence to contribute significantly to the polarization content.  Unlike the leading-order factors, these higher-order
multipoles sometimes treat the component masses \emph{asymmetrically}.  As a result, the waveform and polarization
content near the orbital plane has additional modulations at the orbital period. 
For example, for a nonprecessing
binary with both conjugate and symmetric emission in all $l=2$ modes, the Weyl scalar has the form
\begin{align}
r \WeylScalar(\hat{n},t) = \sum_{m=1}^2 a_m(t) [e^{i  \Phi_m(t)} \Y{-2}_{2,m}(\hat{n}) + e^{-i  \Phi_m(t)} \Y{-2}_{2,-m}(\hat{n})]
\end{align}
As instantaneous quadrupole emission almost always dominates, the polarization amplitude $z_\psi$  can be  approximated by an
oscillating but
nearly-constant correction to the 
leading order term [Eq. (\ref{eq:z:ExampleNonprecessingExtreme})]:   %
\begin{eqnarray}
\label{eq:z:ExampleNonprecessing:WithLeadingOrder}
z_\psi &\simeq &e^{i 4\Psi_L}   \frac{(1-\cos \theta)^2}{(1+\cos \theta)^2} \nonumber
  \\ &\times&  \left(1 + \frac{4  a_1 e^{-i (\phi-\Phi_1+\Phi_2)} \cot\theta}{a_2} + \ldots \right)
\end{eqnarray}
In short, for orientations where higher harmonics contribute significantly, the polarization amplitude $z_\psi$ should
oscillate, with the peak-to-trough amplitude in one-to-one  relation to the magnitude of these harmonics.
To summarize, any gravitational wave signal can be decomposed into two polarizations $\WeylScalar_{R,L}$.    Using the 
complex amplitude $z=\WeylScalar_L^*/\WeylScalar_R$ to characterize the relative proportion and phasing of each signal,
we find that polarization content encodes information about the merging binary on at least two scales.    On the one hand,  on
long precession timescales, the polarization amplitude tells us how the preferred emission direction evolves; see, e.g.,
Eq. (\ref{eq:z:ExampleReconstruct:SlowPrecessing}).   On these scales, the complex amplitude can vary significantly.  On
the other hand, on the orbital and eventually merger timescale, rapid small oscillations in the polarization amplitude
directly measure the relative proportion of  $(2,\pm 1)$ modes.  These modes are produced naturally in
asymmetric or eccentric binaries along certain lines of sight.   
Finally, in our analysis we have for simplicity assumed symmetric emission in the $(2,\pm 2)$ modes.  However, when expressed in a
corotating frame \cite{gwastro-mergers-nr-Alignment-ROS-Methods}, many of our simulations emit
asymmetrically, with one mode preferentially larger than another during the merger and ringdown phase.   Without
foreknowledge of this bias, the simple estimates used above would, if inverted, recover the incorrect inclination of the
preferred emission direction relative to the line of sight.  
In other words, while the complex polarization amplitude $z_\psi$  manifestly encodes information about the preferred
orientation's evolution relative to the line of sight and about the relative significance of higher modes, we cannot
unambiguously interpret this information without the assistance of a large catalog of candidate waveforms.  
For the purposes of this paper, we will treat $z_\psi(t)$ itself as a phenomenological observable.

\subsection{Line of sight diagnostics: Three examples }
\begin{figure}
\includegraphics[width=\columnwidth]{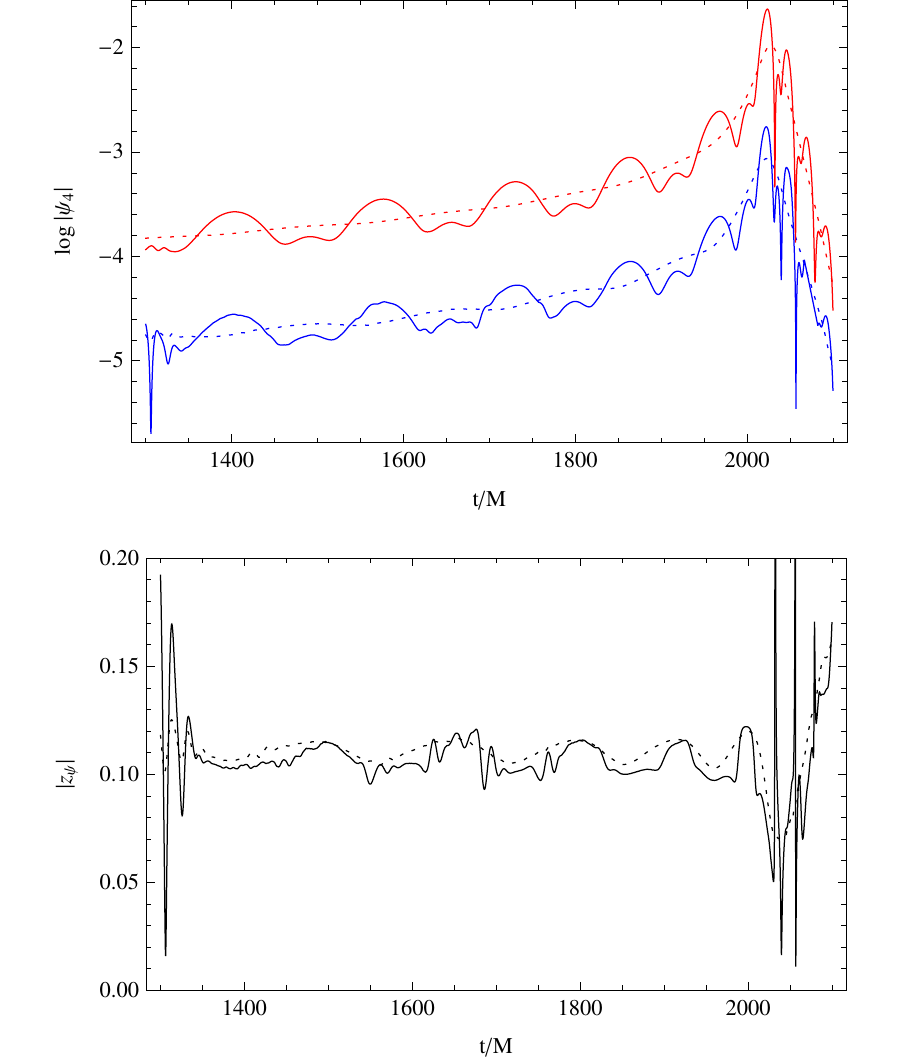}
\caption{\label{fig:PolarizationVersusTime:Fiducial:Nonprecessing}\textbf{Polarization imprints on signal 1:
    Nonprecessing}:  For a nonspinning  $q=4$ binary (Tq(0,4,0)), two measures of the polarization content along
  $(\theta,\phi)=(60^\circ,205^\circ)$.  Results are shown including all angular modes $\WeylScalar_{lm}$ with $l\le 4$ (solid) and just
  retaining terms with $l=2$ (dotted).  %
 \emph{Top panel}: The line of sight right and left-handed amplitudes $|\WeylScalar_{R,L}|$ (red R, blue L).  For this
 and almost all other lines of sight, a single helicity dominates for all time. %
For comparison, the solid black line shows $|\WeylScalar| = |\WeylScalar_R+\WeylScalar_L|$.  
\emph{Bottom panel}: $| z_\psi| =| \WeylScalar_L^*/\WeylScalar_R|$ vesus time.  Though one helicity
dominates, the polarization content oscillates
significantly at the orbital period.  These oscillations are proportional to the ``higher-mode'' content ($2,\pm 1$) and
therefore measure the mass ratio
or residual eccentricity.
}
\end{figure}

\begin{figure}
\includegraphics[width=\columnwidth]{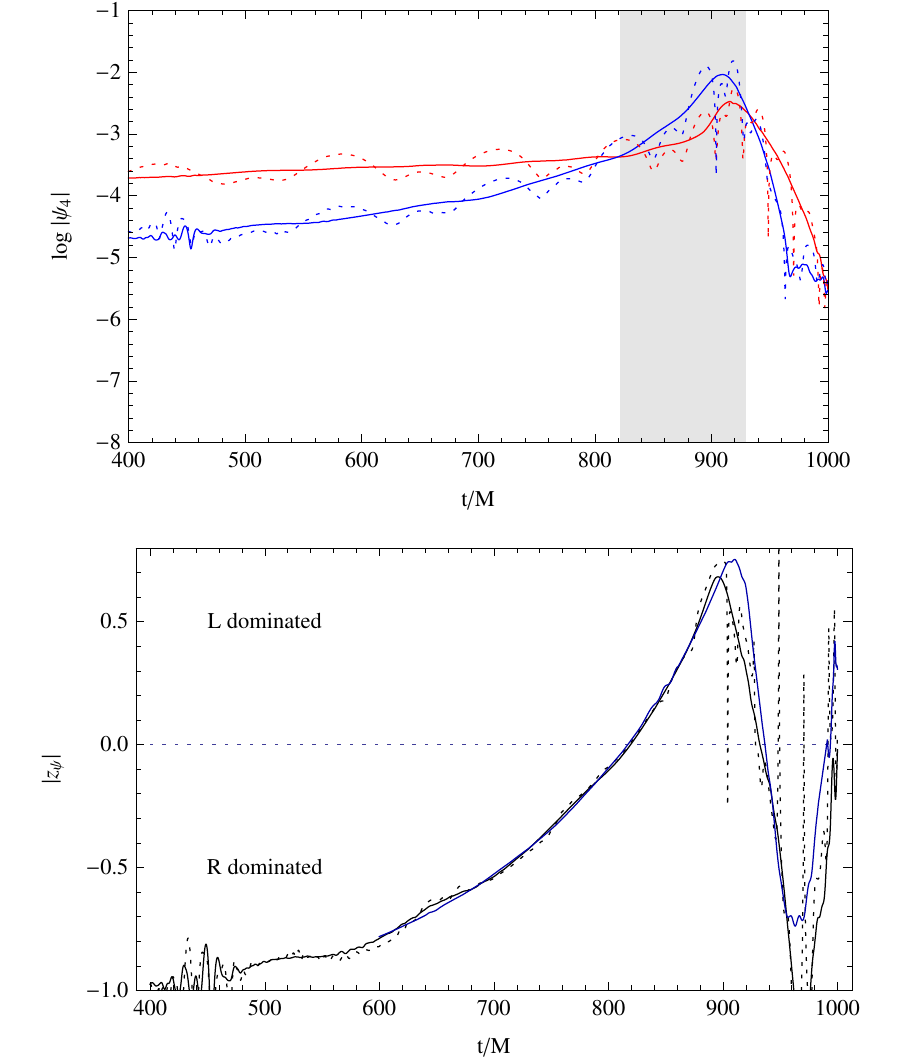}
\caption{\label{fig:PolarizationVersusTime:Fiducial}\textbf{Polarization imprints on signal 2: Precessing}:  For a
  precessing binary (Sq(4,0.6,90,9)), the  polarization carries a strong imprint from the relative orientation of the preferred emission
  direction relative to the line of sight.  For the precessing $q=4$ binary described  in the text and
  \cite{gwastro-mergers-nr-Alignment-ROS-IsJEnough},  a decomposition of the gravitational wave signal along a
  ``generic'' orientation [$(\theta,\phi)=(60^\circ,205^\circ)$] is shown.  Results are shown using just $l=2$ (solid)
  and all modes $l\le 4$ (dotted), \emph{reversing} the convention of Figure \ref{fig:PolarizationVersusTime:Fiducial:Nonprecessing}.
\emph{Top panel}:  The line of sight right and left-handed amplitudes $|\WeylScalar_{R,L}|$ (red R, blue L) using all
modes (dotted) and just $l=2$ (solid).  
The gray shaded box shows the interval where $\hat{V}\cdot \hat{n}\le 0$.  
\emph{Bottom panel}: Comparison of $|z_\psi(t)|$ extracted along this line of sight (black solid [$l=2$] and dotted curves
     [$l\le 4$]) with the leading-order
estimate $\tilde{z}_{\hat{V}}$ provided by Eq. (\ref{eq:z:ExampleReconstruct:SlowPrecessing}) and the preferred
orientation  $\hat{V}$ selected by $\avL_t$ (blue).
}
\end{figure}

\begin{figure*}
\includegraphics[width=0.94\textwidth]{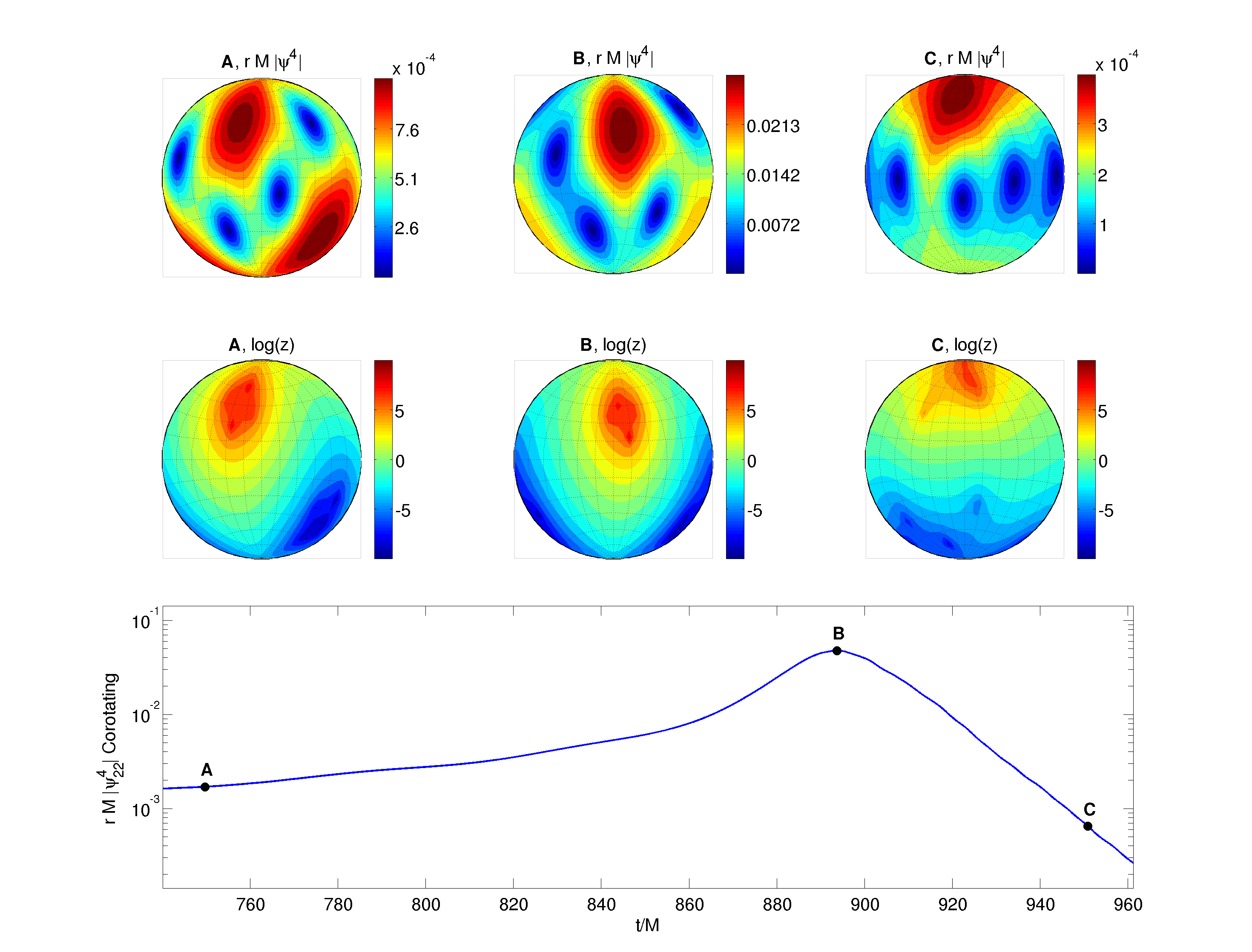}
\caption{\label{fig:PolarizationAtTime:Lionel}\textbf{Evolution of amplitude, polarization at merger}:
 Three snapshots
  of $|\WeylScalar|$ (top panel)  for
  the Sq(4,0.6,90,9) simulation, bracketing the time of peak amplitude and 
  demonstrating that the  polarization content changes significantly during  the merger event.  
For aesthetic reasons, we only show the contributions from all  $l=2$ modes. 
The top panels show the relative scale, with red indicating the largest $|\WeylScalar|$ at that time.  For comparison,
the bottom panel panel illustrates when these snapshots occur, using a plot of   $|\WeylScalar_{22}|$ versus time.
}
\end{figure*}

To illustrate the  power of this decomposition,  Figures
\ref{fig:PolarizationVersusTime:Fiducial:Nonprecessing}, \ref{fig:PolarizationVersusTime:Fiducial}, and 
  \ref{fig:PolarizationAtTime:Lionel} apply it to nonprecessing and
 precessing $q=4$ binaries.

Figure \ref{fig:PolarizationVersusTime:Fiducial:Nonprecessing} demonstrates our polarization decomposition with  a nonprecessing
$q=4$ binary, extracting the Weyl scalar along a ``generic'' direction [$(\theta,\phi)=(60^\circ,205^\circ)$].    For a
nonprecessing binary, the gravitational wave signal is dominated by quadrupole emission from the $(2,\pm 2)$ modes.   To an excellent approximation, the two polarizations $\WeylScalar_{R,L}$ (red and
blue curves in this figure) are proportional to one or the other of these modes, respectively, with the proportionality
constants set geometrically by spin-weighted harmonics [Eq. (\ref{eq:Psi4ToyModel:22Only})] and hence the inclination.    
However, this line of sight is close enough to the orbital plane that the $(2,\pm 1)$ mode can
contribute significantly.    As demonstrated with the dotted line in the bottom panel of Figure
\ref{fig:PolarizationVersusTime:Fiducial:Nonprecessing}, these modes beat against the leading-order quadrupole,
causing the complex polarization $z_\psi$ to oscillate by roughly ten percent during the inspiral at roughly the orbital period.
Higher-order angular modes $l>2$ also beat against the leading-order quadrupole.  Including these terms in the  the sum
$\sum_{lm}\WeylScalar_{lm}\Y{-2}_{l,m}$ produces richer time dependence along each line of sight, in both polarizations   (solid curves in the
top panel of Figure  \ref{fig:PolarizationVersusTime:Fiducial:Nonprecessing}).  Nonetheless, the ratio of the two
polarizations ($z_\psi$) still behaves like the $l=2$ result.

Precessing binaries, by contrast, generally exhibit dramatic changes in polarization.  Figure
\ref{fig:PolarizationVersusTime:Fiducial} shows features of the precessing $M_1/M_2\equiv q=4$ binary started with $\vec{a}_1=0.6
\hat{x} = - a_2$ at a coordinate separation $d=9M$, as used in \cite{gwastro-mergers-nr-Alignment-ROS-IsJEnough}.  
As seen in the top panel, the balance between the two polarizations changes significantly: though R-handed emission
(red) usually dominates, for a short epoch L-handed emission is stronger in this direction (blue).  
Critically, we can identify when this transition occurs by comparing our line of sight with the preferred orientation
$\hat{V}(t)$ extracted from $\avL_t$
\cite{gwastro-mergers-nr-Alignment-ROS-IsJEnough,gwastro-mergers-nr-Alignment-ROS-Methods}.  
Following the discussion above, we expect and our calculations confirm (shaded region) that the two polarizations have equal amplitudes  (i.e., $|z_\psi|=1$) when
$\hat{V}\cdot \hat{n}=0$, with R-handed emission dominating when $\hat{V}\cdot \hat{n}>0$ and L-handed emission
dominating when $\hat{V}\cdot \hat{n}<0$.  
More precisely, we can estimate both the magnitude and phase of $z$ surprisingly reliably by combining
Eq. (\ref{eq:z:ExampleReconstruct:SlowPrecessing}) with the orientation
$\hat{V}(t)$.   For example, the bottom panel of Figure \ref{fig:PolarizationVersusTime:Fiducial}
shows that the simulated polarization amplitude $|z_\psi|$ is extremely close to the estimate $|\tilde{z}_V|$.   Small fluctuations
around the leading-order prediction at the orbital period should be due to residual eccentricity.  

While we have emphasized  $l=2$ modes in the discussion above, the inclusion of higher-order angular modes $l\ge 2$ into
$\WeylScalar$ produces nearly no change to $|z_\psi|$.    In this example, our naive expression relating $z_\psi$ to a preferred direction continues
to apply, even when these harmonics produce dramatic fluctuations in the line-of-sight amplitude.
Judging from this example, similar calculations for the phase of $z_\psi$, and their repeated success for all lines of sight
and simulations considered, we believe $z_\psi$ is primarily determined by the orientation of $\hat{n}$ relative to the time-dependent
preferred orientation  $\hat{V}$ and can usually be well-approximated by  Eq. (\ref{eq:z:ExampleReconstruct:SlowPrecessing}).

As described in subsequent sections, our calculations suggest the  preferred orientation $\hat{V}$ evolves significantly
and rapidly during merger.   
Equivalently,  the polarization content -- the distribution of lines of sight dominated by left versus right handed
emission --  changes significantly at
the merger event.  
 As seen
 in Figure \ref{fig:PolarizationAtTime:Lionel}, 
immediately before and after the merger
event, the two points corresponding to predominantly left- or right-handed emission change noticeably.
This interval corresponds to the merger phase itself.   As described in the next section, we suspect this rapid, global change in polarization content may
reflect features of the strong-field merger event itself.

\subsection{Waveforms along other fixed directions}
\begin{figure}
\includegraphics[width=\columnwidth]{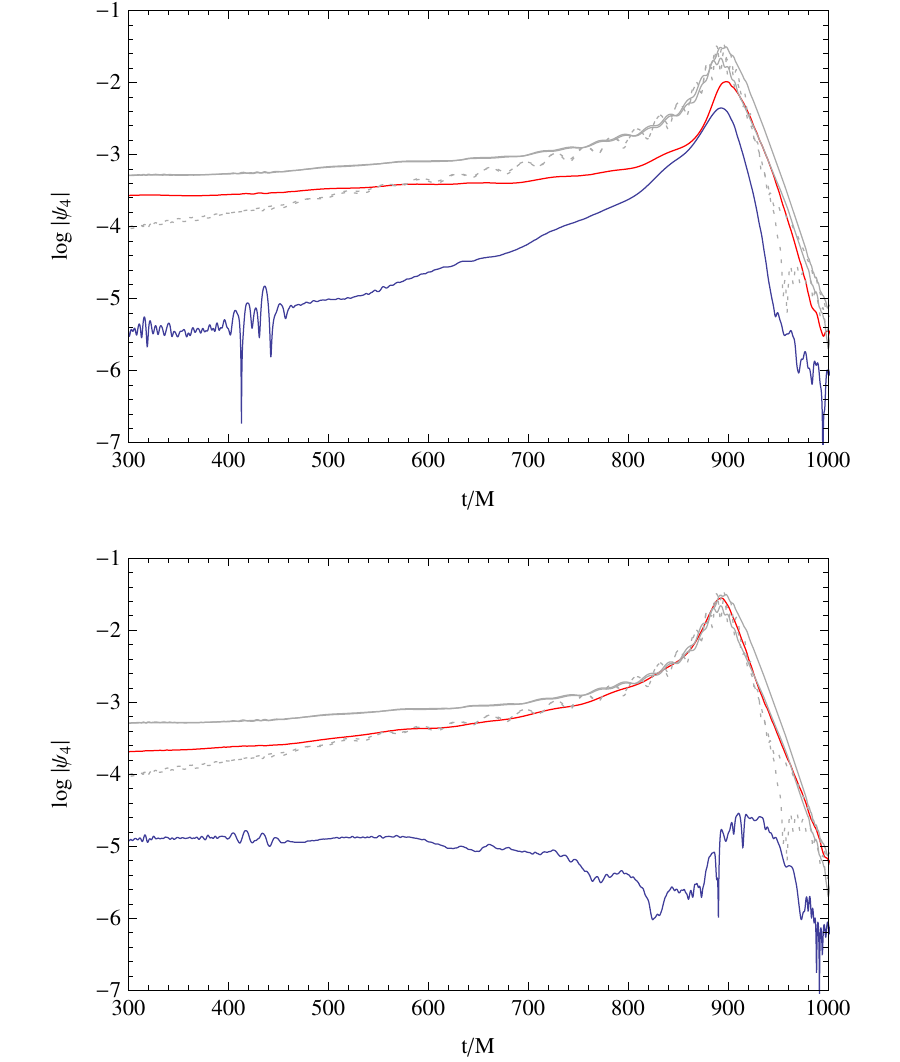}
\caption{\label{fig:PolarizationsVersusTime:OtherOrientations}\textbf{Polarization content along fixed special directions}: The right- and left-handed Weyl scalar amplitude
  $|\WeylScalar_{R,L}|$ (red, blue) versus time, for the gravitational wave signal extracted along several preferred
  orientations, for   Sq(4,0.6,90,9).   For comparison, the gray lines show the simulation-frame $(2,\pm 2)$ (solid gray) and $(2,\pm 1)$
  modes (dotted gray).   For aesthetic reasons, we only show the contributions from the $l=2$ modes. 
\emph{Top panel}:  Initial total angular momentum.  
\emph{Bottom panel}: $\hat{V}$ at the time of peak emission.  From the ratio of left to right (blue to red), the
direction identified by  $\hat{V}$ corresponds to nearly circular polarization near the epoch of peak emission.  This
direction is significantly offset from $\hat{J}$.
}
\end{figure}

For reference, in Figure \ref{fig:PolarizationsVersusTime:OtherOrientations} we show the polarization content for two
other preferred orientations: the initial ($\simeq $ final) total angular momentum direction $\hat{J}$ (top panel) and the preferred
orientation $\hat{V}$ evaluated at the time of peak emission (bottom panel).  
In the first case, one polarization is vastly larger than the other at early times; during the merger, however, both
polarizations become significant.   Similar results are found when extracting along $\hat{z}$, the initial orbital
angular momentum.
In the second case, both polarizations are comparatively large early on.  During the merger epoch, however, only one
polarization dominates.
Generally speaking, when adopting a fixed frame one can choose to simplify some narrow epoch of the waveform by reducing
the other polarization.   For any time, frequency, or mass range, a generalization of $\avL_t$ can be constructed to
determine what orientation would be suitable.   However, in general no one orientation works for all time.

Finally, we emphasize that we have been able to accurately estimate the polarization content using the time-dependent preferred orientation
$\hat{V}$.  For our simulations, this orientation differs substantially from $\hat{L}$ at all times.  Based on this
performance, we anticipate that corotating-frame waveforms along $\hat{V}$ will be substantially simpler than any analog
extracted along  $\hat{L}$.

\section{Simulations II: Trends and variations }
\label{sec:Simulations2}
From the diagnostics above, we anticipate simulations are best and most naturally characterized by (a) the modal waveforms
$\WeylScalar_{lm}$  in a
corotating frame and (b) the evolution of our preferred orientation with time.  In addition, to simplify the translation
between time and frequency domain, we will also use  (c) the overall orientation-averged signal power $\bar{\rho}$.
In this section we briefly report on salient ways these three features change with spin and mass ratio.

\subsection{Polarization bias}
\begin{figure}
\includegraphics[width=\columnwidth]{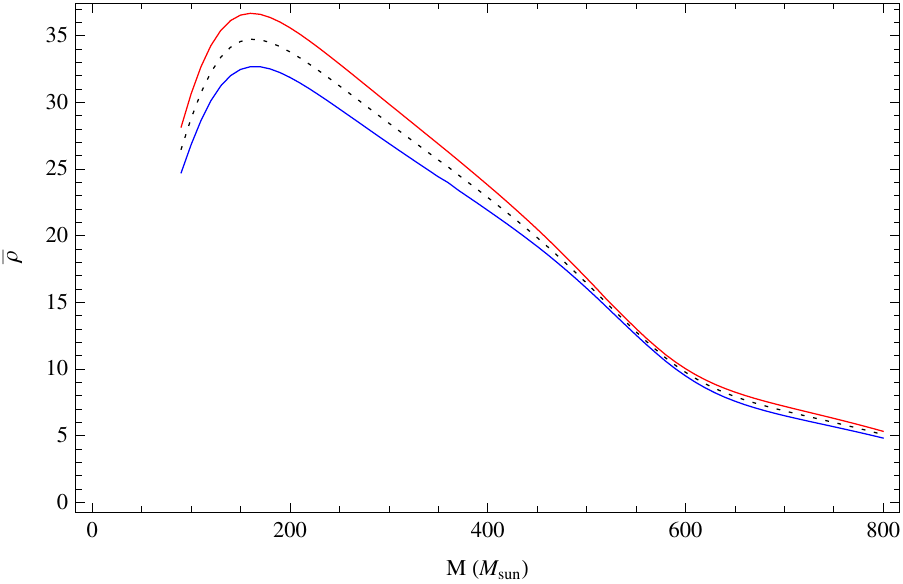}
\caption{\label{fig:Prototype:ChiralBias}\textbf{Merger signals are polarized overall}: Depending on the spins, binary black holes can emit substantially more L
  or R handed power. 
As an example,  a comparison of $\bar{\rho}_R$ (red) and $\bar{\rho}_L$ (blue) to
$\bar{\rho}/\sqrt{2}$ (dotted) for 
the Tq(2,0.6,45) simulation.  In this expression we adopt a fiducial initial LIGO design
noise curve.
}
\end{figure}

Generic precessing binaries exhibit a \emph{polarization bias}: at any given instant, the binary is radiating more of
one handedness than another.   During the inspiral, the balance between L and R oscillates.  At merger, the balance
fixes, preferring one handedness, with the choice depending on the spin-orbit configuration just prior to merger.  This asymmetry  produces large kicks \cite{2008PhRvD..77l4047B}, with
a significant component perpendicular to the orbital plane; see also Healy et al (in prep) and cf.  \cite{2010CQGra..27k4006L}.    

For the purposes of this paper, this asymmetry complicates our interpretation of the preferred direction.  As with
nonprecessing binaries, we use the ratio of left to right-handed power to  estimate an (instantaneous)
\emph{inclination}.  The  expressions that we invert for $\hat{O}\cdot \hat{n}$
     [e.g., Eq. (\ref{eq:z:ExampleReconstruct:SlowPrecessing})] \emph{assume} an equal amount of left and right handed power.
More refined estimates that relate the line of sight to the left/right ratio are required when a strong bias towards one
or the other handedess occurs.

\begin{figure}
\includegraphics[width=\columnwidth]{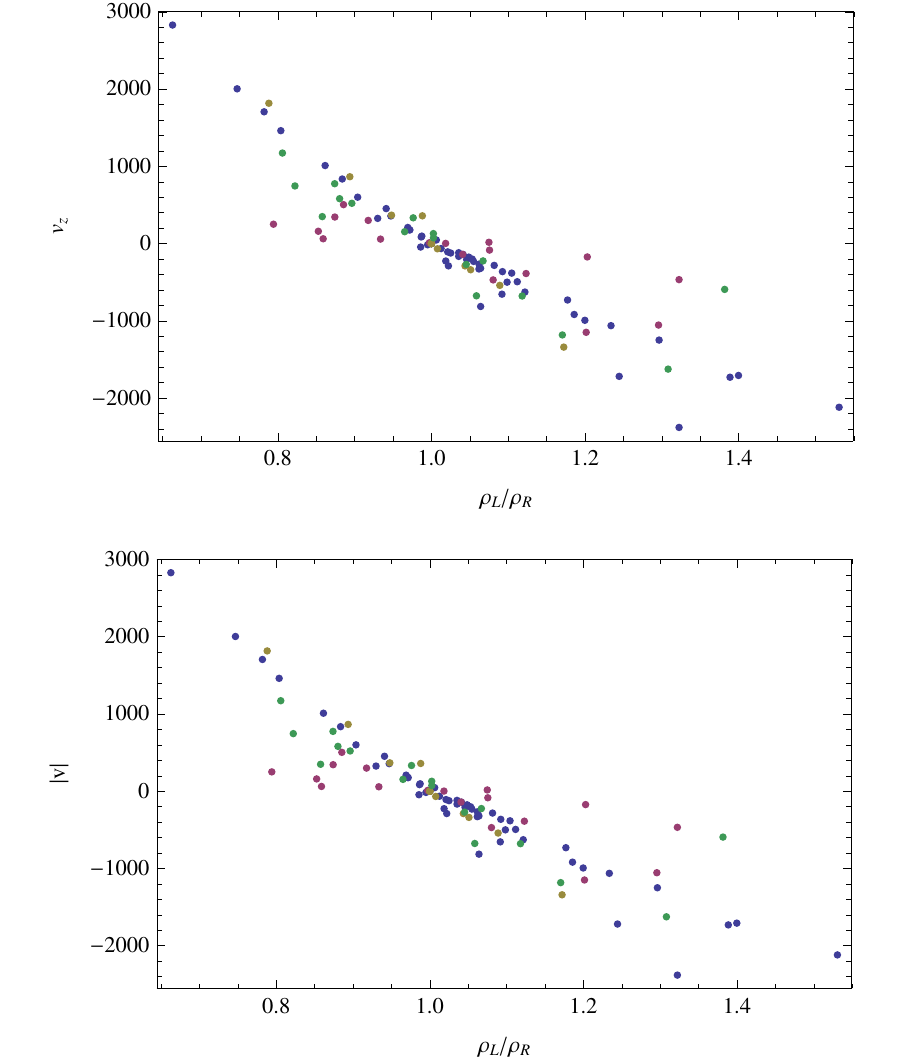}
\caption{\label{fig:KicksVersusAsymmetry}\textbf{Large  L/R asymmetries and large kicks}: Scatter plot of the recoil kick velocity in the $\hat{z}$
  direction (top panel) and overall (bottom panel) versus $\bar{\rho}_L/\bar{\rho}_R$.  Large kicks in the
  $+\hat{z}$ direction correlate with predominantly L-handed emission.    Only a comparatively small amount of
    asymmetry between L and R during merger is required to produce the largest known kicks  \cite{2008PhRvD..77l4047B}.
In this figure, colors correspond to different simulation sets: S (blue), Sq (red), T (yellow), Tq (green).
For illustrating the correlation between R vs L bias and kick magnitude only: kick data is computed using the simulation
series alone at a single extraction radius, without correcting  for the early inspiral. 
}
\end{figure}

To assess whether more complicated expressions would be required in parameter estimation,\footnote{We illustrate polarization
  asymmetry using  a detection-weighted
  diagnostic  to demonstrate that  the bias is \emph{detectable}.   Alternatively and in a detector-agnostic way,
  the polarization asymmetry also shows up clearly in    $(\int d\Omega |\psi_R(\hat{n})|^2)/(\int d\Omega
  |\psi_L(\hat{n})|^2)$, where the numerator and denominator are evaluated at each time or each frequency.} we define
polarized analogs of the orientation-averaged signal amplitude $\bar{\rho}$ \cite{gwastro-spins-rangefit2010}, starting
from $\rho^2$ [Eq. (\ref{eq:def:rho})]:
\begin{eqnarray}
\bar{\rho}^2 &\equiv& \oint \frac{d\Omega}{4\pi} \rho^2(\hat{n}) 
=\oint \frac{d\Omega}{4\pi} 2 \frac{df}{S_h}\frac{|\WeylScalarFourier(f,\hat{n})|^2}{(2\pi f)^4} \\
\bar{\rho}_R^2 &\equiv & \oint \frac{d\Omega}{4\pi} 2 \int_0^\infty \frac{df}{S_h} \frac{|\WeylScalarFourier(f,\hat{n})|^2}{(2\pi
  f)^4}\\
\bar{\rho}_L^2 &\equiv & \oint \frac{d\Omega}{4\pi} 2 \int_{-\infty}^0 \frac{df}{S_h} \frac{|\WeylScalarFourier(f,\hat{n})|^2}{(2\pi f)^4}
\end{eqnarray}
Any given binary has directions where one polarization dominates (e.g, along $\hat{L}$ for a nonprecessing binary).
Summing over all orientations, however, nonprecessing binaries emit symmetrically, with matching amounts of right
(R) and left (L)
handed emission along mirror-symmetry-related lines of sight.   With equal amounts of signal power, nonprecessing
binaries must have  $\bar{\rho}_L = \bar{\rho}_R=\bar{\rho}/\sqrt{2}$.
By contrast, precessing binaries have no symmetry that enforces symmetric emission;  particularly in narrow epochs or
frequency intervals selected by
an outside observer or gravitational wave detector, one handedness (R or L) can dominate.  
As Figure \ref{fig:Prototype:ChiralBias} demonstrates with a specific example and Figure \ref{fig:KicksVersusAsymmetry}
with an ensemble, generally a single polarization does dominate during merger.  
The dominant polarization depends sensitively on the spins, particularly when $\vec{S}_1,\vec{S}_2$ are nearly antiparallel and in the
orbital plane.

Our calculations suggest that $\simeq 30\%$ changes in the overall amplitude ($\bar{\rho}$ or $\psi_4$) are not uncommon
during merger.  The \emph{relative} amplitude of 
left to right handed radiation ($\rho_R/\rho_L$  or $z$), however,  changes by orders of magnitude due to small changes in inclination; see
Eq. (\ref{eq:z:ExampleReconstruct:SlowPrecessing}) and Figure \ref{fig:PolarizationVersusTime:Fiducial}.  As a result,
the typical polarization bias introduces a fairly small systematic error into any procedure to reconstruct the evolution
of $\hat{V}$ [e.g., into the inverse of
Eq. (\ref{eq:z:ExampleReconstruct:SlowPrecessing}) for the inclination as a function of time].

\subsection{Corotating waveforms,  chirality}

The corotating-frame waveforms will be described in detail in a subsequent publication \cite{gwastro-mergers-nr-Alignment-ROS-CorotatingWaveforms}.  For the purposes of this paper,
we will employ only two key features.  
First and foremost, like their nonprecessing analogs, the corotating frames are \emph{chiral}: modes with $m>0$ have
frequency content only for $f>0$ and vice-versa.    We can quantify how precisely we are confident that the modes are
chiral in a way that is relevant to data analysis.  For example, we can decompose the contribution of each
\emph{corotating} mode to \emph{ficticious}\footnote{Ficticious since they are associated with a corotating frame.}   ``orientation-averaged
signal-to-noise ratios'' $\bar{\rho}_{R,L}$ associated with each handedness:
\begin{eqnarray}
\bar{\rho}_{lm,R,corot}^2&\equiv& \frac{1}{4\pi}  2\int_{0}^{\infty} \frac{df}{S_h} \frac{|\psi_{4,lm}^{corot}(f)|^2}{(2\pi f)^4}\\
\bar{\rho}_{lm,L,corot}^2&\equiv& \frac{1}{4\pi}  2\int_{-\infty}^{0} \frac{df}{S_h} \frac{|\psi_{4,lm}^{corot}(f)|^2}{(2\pi f)^4}
\end{eqnarray}
Applying these expressions, we find $\bar{\rho}_{lm,L}/\bar{\rho}_{lm} \simeq 0$ for $m>0$ : modes that should have
positive helicity ($m>0$) have little negative-frequency power.
As a corollary, the corotating-frame $(2,2)$ and $(2,-2)$ modes are \emph{nearly orthogonal} -- as, more generally, are the subspaces spanned by modes with $m>0$ and  $m<0$.
We will use the orthogonality of the corotating-frame $(2,2)$ and $(2,-2)$ modes to model how precisely we can measure
the orientation of the corotating frame, both on average and as function of time.

\subsection{Preferred direction precesses}
\label{sec:sub:Precess}

\begin{figure}
\includegraphics[width=\columnwidth]{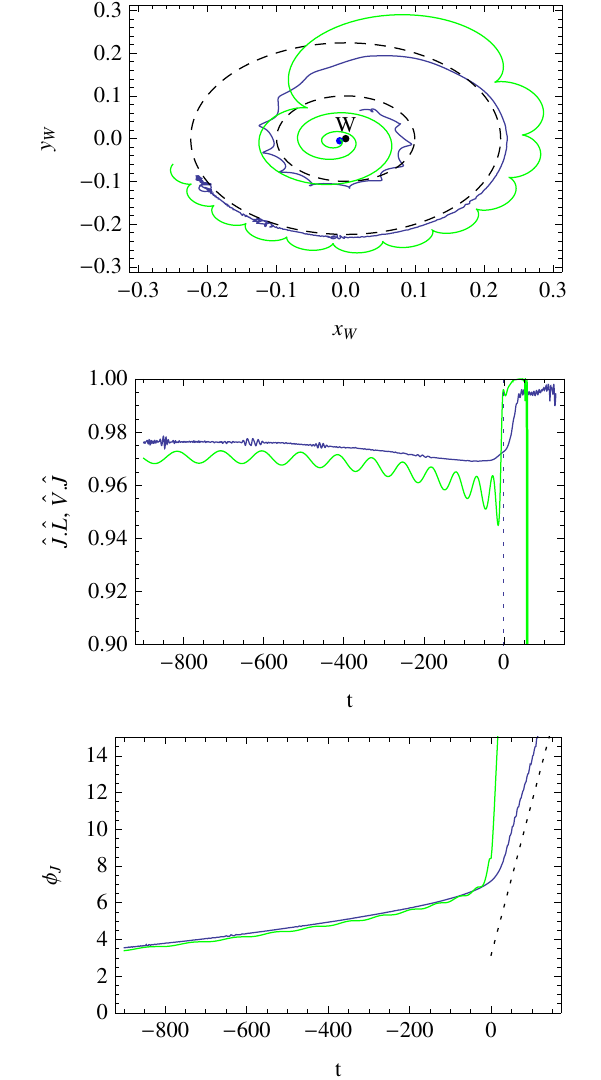}
\caption{\label{fig:PrecessionOfV:Example}\textbf{Precession of preferred orientation: Example}: Demonstration that the
  preferred orientation precesses around $\hat{W}\simeq \hat{J}$
for $q=2,  d=10M, a_1=0.6 \hat{x},a_2=0.6 \hat{z}$.  
\emph{Top panel}: Two-dimensional path of  $V$ (blue) and $\hat{L}$ (green) around a proposed value of $W$ (point at origin), as seen in a plane perpendicular to
$\hat{W}$.  The two dashed curves are circles of constant opening angle.  For comparison, a blue point indicates the
final angular momentum direction $\hat{J}$ in this frame. 
\emph{Center panel}:  $\hat{J}\cdot \hat{V}$ (blue) and $\hat{J}\cdot\hat{L}$ (green), showing both $\hat{L}$ and $\hat{V}$ precess along
similar cones, with a nearly constant opening angle prior to merger.   In this figure, merger occurs
at $t\simeq 0$.
\emph{Bottom panel}:  Plot of $\phi_{VW}$ before and after merger, demonstrating an abrupt change in the precession
frequency at the merger event.  
For comparison, the green curve shows $\phi_{L}$, the precession phase extracted from $L$ around $J$.
For this system, at late times the ``precession rate''
$\Omega_{VW}\equiv \partial_t \phi_W \simeq 1/12 M$ is still an order of magnitude smaller than $\partial_t \Phi$, the
``carrier frequency'' set by the $(2,2)$ and $(2,-2)$ modes. %
}
\end{figure}

For the simulations and time intervals we have simulated, the preferred orientation $\hat{V}$ evolves as if precessing
along a nearly-constant cone, centered along some axis $\hat{W}$.    Figure \ref{fig:PrecessionOfV:Example} provides a
concrete example.    In almost all cases we find $\hat{W}\simeq
\hat{J}_{\rm final}$
empirically: $\hat{V}$ precesses around the total angular momentum.    For  this section only, however,
we  allow $\hat{W}$ to take arbitrary values.

Given the orientation and an arbitrary frame $\hat{x}_W,\hat{y}_W$ defined perpendicular to the constant vector $\hat{W}$, we define the
precession phase $\phi_W$ and precession frequency $\Omega_{VW}$ via
\begin{subequations}
\label{eq:def:OmegaVW}
\begin{eqnarray}
e^{i \phi_{VW}}  &=& \frac{(\hat{x}_W+i \hat{y}_W)\cdot \hat{V}}{|(\hat{x}_W+i \hat{y}_W)\cdot \hat{V}|} \\
\Omega_{VW} &=& \partial_t \phi_{VW}
\end{eqnarray}\end{subequations}
and make similar definitions for $\hat{L}$ around $\hat{J}$.   
 During the inspiral, the two vectors $\hat{V}$ and $\hat{L}$ nearly coincide, precessing around a similar center at a
 similar rate.   To quantify this similarity, Figure \ref{fig:PrecessionOfV:Example} shows the opening angle of each
 precession cone ($\hat{J}\cdot \hat{L}$ and $\hat{J}\cdot \hat{V}$) and the ``phase'' $\phi_{VJ}$ and $\phi_{LJ}$ of
 the vectors $\hat{L}$ and $\hat{V}$ as they precess around $\hat{J}$.  Prior to merger, both quantities largely agree.

As the binary plunges and merges, the precession cones of both $\hat{L}$ and $\hat{V}$ %
cycle around their
 respective axes ($\hat{J}$ and $\hat{W}$) at a higher rate.   By comparison to $\hat{L}$, however, the preferred axis
 $\hat{V}$ has a much smaller precession rate and persists in precession cycles long after merger.
Furthermore,  contrary to our intuition, the opening angle of the precession cone for $\hat{V}$ (i.e.,
$\arccos(\hat{W}\cdot \hat{V})$) rarely decays significantly after merger.
This precession cycle of $\hat{V}$ is robust: both for this simulation and all others tested, all extraction radii and
(when available)  all resolutions show quantitatively similar features.

For each simulation, we estimated $\hat{W}$ and $\phi_W$ by fitting a fixed precession cone to the trajectory of
$\hat{V}$.    Specifically, we adopt as  $\hat{W}$
the direction that mimimizes  the rms difference between $\hat{V}(t)\cdot \hat{W} $ and its time average.  %
This choice corresponds to the assumption that the precession cone swept out by $\hat{V}$ should have a constant opening
angle.   When estimating $\hat{W}$, we separately fit data \emph{prior to merger} and \emph{after merger}.  In almost
all cases we found $\hat{W}\simeq \hat{J}$ and $\Omega_{VJ}\simeq \Omega_{VW}$.
To further demonstrate the close correspondence between these two directions, Figure
\ref{fig:Simulations:VPrecession:Frequency} shows \emph{both} angular frequencies: points appear at the median
value $(\Omega_{VW}+\Omega_{VJ})/2$; the bar has height $\pm (\Omega_{VW}-\Omega_{VJ})/2$.  
Except for a handful of cases, mostly associated with nearly-aligned spins, these two expressions agree.  
We conservatively adopt the difference $(\Omega_{VW}-\Omega_{VJ})/2$ as an estimate of our systematic error.  
[Other sources of systematic error, such as extraction radius, have a smaller effect on the recovered post-merger
  precession rate.]

\subsection{Post-merger precession 1: Results}
In our simulations, if the binary precesses \emph{prior} to merger, our preferred direction $\hat{V}$ continues to
precess \emph{after} merger, never converging towards $\hat{J}$.     Precession is ubiquitous (i.e., whenever $\hat{L}$
and $\hat{J}$ are initially misaligned); 
we therefore expect ubiquitous  precession-induced modulations at and after merger when asymmetric binaries merge with
randomly-oriented  spins.
As described in a forthcoming companion publication \cite{gwastro-mergers-nr-Alignment-ROS-CorotatingWaveforms}, we have performed extensive resolution tests and remain confident that the
post-merger oscillations we observe are resolved. 
Table \ref{tab:Simulations:VPrecession} provides an estimate of the post-merger   ``precession frequencies'' $\Omega_{VJ}$ for each
of our simulations.  

\begin{figure}
\includegraphics[width=\columnwidth]{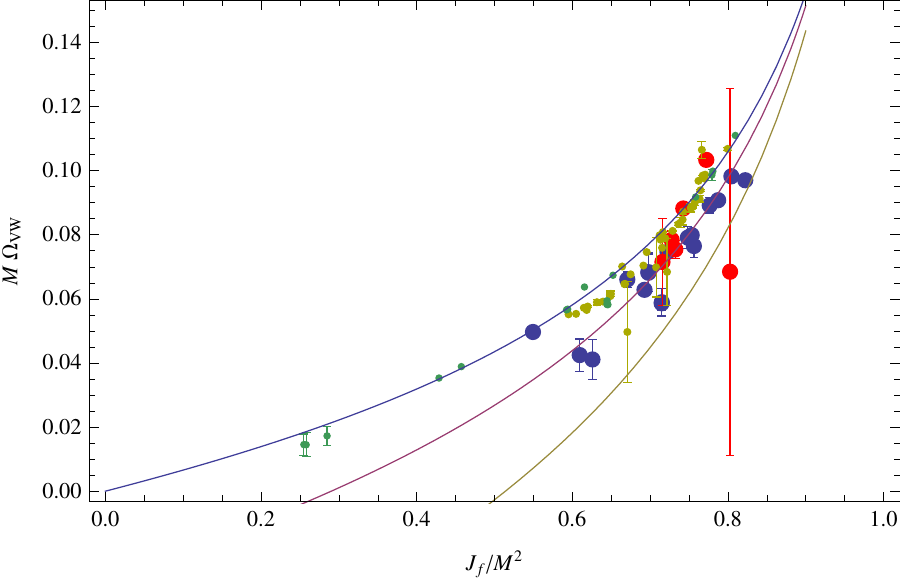}
\caption{\label{fig:Simulations:VPrecession:Frequency}\textbf{Post-merger precession frequency}: For each of the
  simulations listed in  Table \ref{tab:Simulations:VPrecession}, a scatterplot of their ``post-merger precession
  frequency'' $M_f \Omega_{VW}$ [Eq. (\ref{eq:def:OmegaVW})] against the final black hole's spin magnitude $J_f/M_f^2$. 
 Colors indicate four different classes of simulation: Tq (blue), T (red), Sq (green) and S (yellow).
For comparison, the solid lines show the quasinormal mode frequency differences $\omega_{122}-\omega_{121}$
(blue), $\omega_{222}-\omega_{121}$ (red), and $\omega_{322}-\omega_{121}$ (yellow),
where $\omega_{nlm}$ is the real part of the quasinormal eigenfrequency  \cite{2009CQGra..26p3001B}.
The bars indicate differences between two methods for estimating $\Omega$; see Section \ref{sec:sub:Precess}.
}
\end{figure}

   Figure \ref{fig:Simulations:VPrecession:Frequency} demonstrates that the post-merger ``precession''
frequencies are principally determined by the final black hole's spin magnitude.  
Because the final black hole can be described as a superposition of quasinormal modes, the precession frequencies are
very nearly  \emph{differences} 
between the $(l,m)=(2,2)$ and $(2,1)$ quasinormal modes of the remnant hole \cite{2009CQGra..26p3001B}. %
 By some change of basis, the long-lived post-merger oscillations must be described as some
perturbation of the final remnant black hole: a (coherent) superposition of multiple quasinormal modes.   
Our calculations suggest precessing mergers coherently excite multiple quasinormal modes of different angular order.   
These quasinormal modes have similar decay timescales ($1/\text{Im}(\omega)$)  but noticeably different characteristic frequencies
($1/\text{Re}(\omega)$).  As a result, though in principle the quasinormal modes are not \emph{exactly} degenerate, in \emph{practice}
just after merger these coherently-excited quasinormal  modes decay at a similar rate but with changing relative phase,
leading to a black hole state that appears to precess. 
To illustrate why our preferred direction precesses,  we  consider a toy model that approximates   features of a
superposition of outgoing quasinormal modes:
\begin{eqnarray}
r \WeylScalar_{con}(t) &\equiv& \sum_{m} a_{2|m|}e^{-i\omega_{lm} t}\Y{-2}_{l,m}
\end{eqnarray}
where $\omega_{lm}$ is the lowest-order complex eigenfrequency for the $(l,m)$ eigenspace, where the eigenmodes are defined
relative to the black hole's  angular momentum direction $\hat{J}$, and where $a_{2|m|}$ are three  parameters, here assumed
real for simplicity.    
In this example, we ignore the difference between the actual angular eigenfunctions of the Kerr-background wave
equations (spin-weighted spheroidal harmonics) and the corresponding eigenfunctions for flat space, $\Y{-2}_{l,m}$.
In the trivial case where only $a_{2\pm 2}$ are nonzero, the preferred orientation remains along the black hole's
angular momentum axis $\hat{J}$.  %
Allowing $a_{2\pm 1}$ to be nonzero but keeping $a_{20}=0$ leads to a preferred orientation $\hat{V}$ that spirals inward
exponentially towards $\hat{J}$, precessing around $\hat{z}$ with a frequency $\Omega_{VJ}=\text{Re}(\omega_{22}-\omega_{21})$ and
shrinking towards $\hat{z}$ at a different rate $\Gamma_{VJ} = \text{Im}(\omega_{22}-\omega_{21})$.
Finally, when all mode amplitudes are nonzero, the preferred orienation can exhibit a wide range of behaviors depending
on the relative mode ratios $a_{21}/a_{22}$ and $a_{20}/a_{22}$, including
exponential decay to $\hat{J}$; modulated precession around $\hat{J}$;  and pathologically complicated behavior in the
presence of degeneracies.  

\begin{table*}
 {\tiny
 \input{tab-mma-SimulationSet.tex}
 }
\caption{\label{tab:Simulations:VPrecession}\textbf{Simulations used}:
{\tiny The first column is a key, encoding
  the family, mass ratio, black hole spin magnitude $|S_1|/M_1^2=|S_2|/M_2^2$ and alignment.   The next  8 columns
  provide specific initial conditions: the initial separation ($r_{start}$), mass ratio $q=M_1/M_2$, and two component spins
  $S_k^2/M^2$ relative to the total initial mass.  The next column provides $\bar{\rho}_L/\bar{\rho}_R$, a measure of
  whether the  merger event preferentially
  radiates R or L-handed emission.   [This quantity is evaluated for $M=200
  M_\odot$ and the initial LIGO design noise curve.]  
  The next column provides $\Omega_{VJ}$, the post-merger ``precession frequency'' of $\hat{V}$ 
  around $\hat{J}$, estimated using $\simeq 50 M$ after peak $(2,2)$ emission; ``unk'' indicates entries where we cannot
  reliably determine it.
  The last three columns provide the simulation duration, the length of the reliable waveform, and the highest
  resolution  $h$ used.  
}}
\end{table*}

In this interpretation, the angle between $\hat{V}$ and $\hat{J}$ reflects the amplitude of the $(2,\pm 1)$ modes.
Conversely, the angle between $\hat{V}$ and $\hat{J}$ is also partially \emph{geometric}, being tied to the spin-orbit
configuration and how the binary evolves during merger.   However,  as is apparent from  Figure \ref{fig:PrecessionOfV:Example}, 
the late-time $(2,1)$ amplitude (i.e., the angle $\hat{V}\cdot \hat{J}$) \emph{evolves} during the strong-field merger event.   
We anticipate that the opening angle will provides  insight into the strong-field merger process.

\subsection{Post-merger precession 2: Broader context }
Our results suggest that the  nonlinear merger event coherently seeds multiple quasinormal modes.  After merger, all
quasinormal modes appear to precess almost coherently, in a frame aligned with the final black hole's spin.  For example, each constant-$l$ subspace seems to precess at a
similar rate.  We defer a detailed discussion of each constant-$l$ subspace to subsequent publications.  

Our results agree with a recently-developed geometric interpretation of Kerr quasinormal modes.   
Using a short-wavelength limit, \citet{gwastro-YangZimmermanEtc-QuasinormalSpectrumAndPrecession2012} recently showed
that  these kinds of quasinormal mode frequency differences are intimately connected to  precession of spherical photon orbits.  
Our calculations suggest this kind of geometric interpretation holds even in the strong field, during
merger.

Because rapid polarization changes occur during the merger epoch,  we anticipate  that these polarization changes encode linear and potentially nonlinear features of strong-field
gravity.    
At a minimum, the precession cycle of $\hat{V}$ persists during and after the merger, with a precession rate tied to the
quasinormal mode frequency distribution of the final hole.  Modulations in the line of sight polarization content
 provide a natural way to identify this characteristic frequency, potentially allowing new experimental tests of general
 relativity.  
Second, the long-term, multimodal coherence we observe suggests that a
short,  nonlinear, and nontrivial process  seeds a coherent
multimodal perturbation of the final black hole.     
In part, of course, the perturbed black hole  simply inherits features from the just-prior-to-merger binary.  In the
geometric point of view,  $\hat{V}\cdot \hat{J} \ne 0 $ prior to merger; in the quasinormal mode view, the $(2,1)$ mode
just prior to plunge partially limits the $(2,1)$ quasinormal mode amplitude during and after merger.    However,  the merger
process partially but \emph{incompletely} sheds this quantity.    
We anticipate that the amplitude of this effect, measured either by  $\hat{V}\cdot\hat{J}$ or the peak $(2,1)$
amplitude in a frame aligned with $\hat{J}$, will provide a useful probe of the strong-field merger process.

\subsection{Recovering the direction using polarization}

For precessing binaries, we have argued that the line-of-sight polarization, as a function of time, determines both the
line of sight and the path of $\hat{V}$, a preferred direction.   We have argued that this process works for all time, both
prior to and after merger.     
In the theorists' paradise -- access to a noise-free line-of-sight signal $\WeylScalar(t)$ -- we have already arrived at this
paper's key result: 
by tracking polarization, we can partially reconstruct  properties of the source binary, even when limited to the short merger epoch.

In practice,  however, gravitational wave detectors have limited sensitivity, with access to only a small time and
frequency interval of the full signal.  Nontheless, as we argue below,  both the polarization bias,
$\bar{\rho}_L/\bar{\rho}_R$, and preferred orientation, $\hat{V}$, can be \emph{measured} by tracking the time-dependent
polarization along a single line of sight.   

\section{Polarization is measurable}
\label{sec:PolarizationImprint}

In the previous section we described how to characterize the polarization content of a gravitational wave signal.  In
this section we demonstrate that this content is experimentally accessible, both on average and to a lesser extent as a
function of time.  We
quantify the amount of polarization difference that experiments can distinguish.  In particular, we show how fairly small
polarization fluctuations can lead to substantial differences between nonprecessing and precessing waveforms.

\subsection{Nonprecessing binaries}
As shown above [Eq. (\ref{eq:z:ExampleNonprecessingExtreme})], for nonprecessing binaries measurements of the
polarization content $z$ are equivalent to constraints on the emission inclination ($\theta$).
While phrased in different coordinates and applied to an entirely different mass regime where semianalytic waveform
models exist, several extensive discussions of parameter estimation including inclination exist in the literature
\cite{1995PhRvD..52..848P,gw-astro-Vallis-Fisher-2007,CutlerFlanagan:1994,2008ApJ...688L..61V,2009CQGra..26k4007R,gwastro-mergers-PE-Aylott-LIGOATest,2011ApJ...739...99N,2011PhRvD..84b2002L}.
For a signal of amplitude $\rho$, the inclination can be measured to order
\[
\Delta (\cos \theta) \simeq 1/\rho
\]
Detailed Fisher matrix and Markov-chain Monte Carlo calculations corroborate this simple estimate 
\cite{1995PhRvD..52..848P,gw-astro-Vallis-Fisher-2007,CutlerFlanagan:1994,2008ApJ...688L..61V,2009CQGra..26k4007R,gwastro-mergers-PE-Aylott-LIGOATest,2011ApJ...739...99N,2011PhRvD..84b2002L}.

For nonprecessing binaries dominated by $l=|m|=2$ emission, the polarization content is equally a measure of the relative amplitude (and phase) of the
$(2,2)$ versus $(2,-2)$ modes; see Eq. (\ref{eq:z:ExampleNonprecessingExtreme}).  Because the  $(2,2)$ and
$(2,-2)$ modes have definite (and opposite) helicity, we can equally interpret inclination measurements and
polarization content constraints as information about
 the relative importance of left- and right-handed emission, \emph{averaged over the whole signal}.  

\subsection{Static and dynamic polarization content accessible}

As demonstrated extensively for nonprecessing signals, gravitational wave observations can constrain the polarization
content of a signal with constant polarization, a ``static'' signal.  By implication, observations can equally well constrain the ``static part'' of a
signal with weakly-time-varying polarization content.   Generalizing,  observations should also be able to constrain
some ``average polarization'' from an arbitrary gravitational wave signal.   How?  Conceptually, gravitational wave
detector networks  can be arranged to be sensitive to only one (circular or linear) polarization at a
time.  By measuring the total power incident in each polarization,  we directly constrain the ``average polarization.''

A detailed analysis of the response of real gravitational wave detector networks to each circular polarization is beyond
the scope of this paper.  Instead, for simplicity and following the philosophy outlined in
\cite{gwastro-mergers-nr-Alignment-ROS-IsJEnough} and near Eq. (\ref{eq:def:rho}), we will idealize
gravitational wave networks as equally sensitive to both linear polarizations.   
By construction, the left- and right-handed components of $\WeylScalar$ are orthogonal.  The net signal amplitude
$\rho^2(\hat{n})$ can be expressed as a sum of left- and right-handed components:
\begin{eqnarray}
\rho_R^2(\hat{n})&\equiv& 2\int_{0}^{\infty} \frac{df}{S_h} \frac{|\psi_4(f,\hat{n})|^2}{(2\pi f)^4}\\
\rho_L^2(\hat{n})&\equiv& 2\int_{-\infty}^{0} \frac{df}{S_h} \frac{|\psi_4(f,\hat{n})|^2}{(2\pi f)^4}
\end{eqnarray}
Observations determine the integrated contributions from both polarizations ($\rho_{R,L}$) independently.  In this
averaged sense, observations can unambiguously determine the polarization content: we can simply measure the (source
mass and detector-dependent) ratio
\begin{eqnarray}
\bar{z}(M)\equiv \rho_L/\rho_R \; .
\end{eqnarray}
For nonprecessing binaries, this average polarization amplitude $\bar{z}$ corresponds precisely to the instantaneous polarization
amplitude given in Eq. (\ref{eq:z:ExampleNonprecessingExtreme}).
Motivated by the accuracy of inclination measurements and fluctuations in each (independent) polarization, we anticipate this quantity can be measured to order $\Delta \ln \bar{z} \simeq 1/\text{min}(\rho_R,\rho_L)$.

For sufficiently strong signals, the Weyl scalar can be reconstructed in the time domain, allowing us to estimate
$\WeylScalar_{R,L}$ and therefore the polarization content at each time.   Nonparametric signal reconstruction
algorithms like \texttt{coherent waveburst} are already employed in attempts to identify and reconstruct signals from
unmodeled sources  \cite{CWaveburst1,2008CQGra..25k4029K}.
At the signal amplitudes expected from the first merger detections, however, fully nonparametric models generally have
too much freedom to tightly constrain the polarization evolution seen in merging black hole binaries.
Nonetheless, tighter constraints should be achievable, to the extent that precession-related timescales remain long compared to other
scales in the signal.

In fact, several post-Newtonian parameter estimation studies have  demonstrated how well these longer
timescales can be constrained and differentiated, using both  ground- and space-based interferometers
\cite{2009CQGra..26k4007R,2011ApJ...739...99N,2011PhRvD..84b2002L}.    Implicitly, the initial conditions of an inspiralling, precessing
binary encode its precession trajectory, albeit in suboptimal coordinates.\footnote{The traditional coordinate system
  for binary parameters adopts the \emph{initial conditions} and \emph{spin vectors}, rather than the geometry of the
  precessing binary as it passes through the sensitive band.  Recent calculations suggest coordinates adapted to the
  center of the band and which phenomenologically encode precession will more transparently represent the available
  information; see, for example, \citet{gw-astro-SpinAlignedLundgren-FragmentA-Theory}.
}   To the extent that  post-Newtonian studies of binary parameter estimation confirm that both spins and directions of merging
binaries can be measured, they also imply that the (time-dependent) precession trajectory can be distinguished. 
Furthermore, post-Newtonian  simulations  suggest that even marginal precession (i.e., even less than one precession
cycle) has a dramatic impact: if precession is included, then precessing signals can appear dramatically different than
their similar nonprecessing counterparts.

\subsection{Polarization-induced mismatch: Loss of amplitude relative to nonprecessing}
\label{sec:sub:DemonstrateLossOfSNR}

To demonstrate the critical impact even weak but \emph{time-dependent} precession has on a merger signal, we remove it.
Specifically, we start with  our
standard  precessing binary [Sq(4,0.6,90,9)], examine all possible emission directions, and determine how well a comparable nonprecessing
analog could fit them.  
 To reduce ambiguity, for our ``nonprecessing'' analog we use the  \emph{corotating-frame}
$(2,\pm 2)$ modes from that same simulation.\footnote{Alternatively, one can estimate how much precession matters by
   comparing a precessing NR signal with a nonprecessing template family, including a full search over all possible
   component masses and emission directions.  This challenge will be addressed in a subsequent publication, using 
   existing and recently developed numerical and theoretical signal models.}   The procedure for constructing these waveforms is described in a companion
 publication \cite{gwastro-mergers-nr-Alignment-ROS-CorotatingWaveforms}.   As noted above, these two corotating modes  are orthogonal, with support
only for $f>0$ or $f<0$ respectively.   Specifically, a nonprecessing analog of our binary would be well-approximated by
any linear combination of the $(2,2)$ and $(2,-2)$ modes.

As in our prior work \cite{gwastro-mergers-nr-Alignment-ROS-Methods,gwastro-mergers-nr-Alignment-ROS-IsJEnough}, for
simplicity we compare two complex waveforms using an inner product that accounts for both polarizations simultaneously:
\begin{eqnarray}
\label{eq:def:Overlap}
(A,B) &\equiv&  \int_{-\infty}^{\infty} 2 \frac{\tilde{A}(f)^* \tilde{B}(f)}{(2\pi f)^4 S_h} \,df
\end{eqnarray}
In this expression we are not maximizing over time or phase.   Using this inner product, we can define two
normalized L and R-handed basis states from the corotating-frame modes:
\begin{eqnarray}
\qmstate{R} = \WeylScalar_{22}(t)/\sqrt{(\WeylScalar_{22},\WeylScalar_{22})} \\
\qmstate{L} = \WeylScalar_{2-2}(t)/\sqrt{(\WeylScalar_{2-2},\WeylScalar_{2-2})}
\end{eqnarray}
Using these orthonormal basis states, we can construct the ``nonprecessing signal'' $\psi_{est}$ that best resembles $\psi(t,\hat{n})$:
\begin{eqnarray}
\qmstate{\psi_{est}} &=& (R,\psi) \qmstate{R} + (L,\psi) \qmstate{L} \equiv {\cal P}\psi
\end{eqnarray}
where ${\cal P}$ is a projection operator to the subspace spanned by $R,L$. 
From the difference between this state and the original state, we can determine how much information is lost by a
nonprecessing approximation for each candidate line of sight:
\begin{eqnarray}
\rho_{rem}^2 &\equiv& (\psi, (1-{\cal P})\psi) = \rho^2 - ( |(R,\psi)|^2 + |(L,\psi)|^2) 
\end{eqnarray}
This expression can be further simplified by expanding $\psi$ in left- and right-handed functions of time along that
line of sight:
\begin{eqnarray}
\qmstate{\psi}&\equiv& \rho_L \qmstate{\hat{\psi}_L} + \rho_R \qmstate{\hat{\psi}_R} =  \rho\qmstate{\hat{\psi}}
\end{eqnarray}
where for clarity we adopt hats to denote normalized signals (e.g.,  $\qmstate{\hat{\psi}}=\qmstate{\psi}/\rho$).  
Substituting this expression  then using orthogonality of L and R handed signals gives an expression for  the SNR lost
in terms of the intrinsic L and R handed amplitudes $\rho_{R,L}$ and the \emph{overlaps} $(R,\hat{\psi}_R)$ and $(L,\hat{\psi}_L)$ between the
two (normalized) line of sight R and L basis signals  and our two R- and and L-handed basis functions:
\begin{subequations}
\label{eq:rhoLost}
\begin{align}
\frac{\rho_{rem}^2}{\rho^2} &=& 1 - \frac{1}{\rho_R^2 + \rho_L^2}[\rho_R^2|(R,\hat{\psi}_R)|^2 + \rho_L^2
  |(L,\hat{\psi}_L)|^2 ]
\end{align}
In the above discussion we  pessimistically perform a point-to-point comparison: we do not allow the $R,L$
waveforms to vary in phase or time, neither relative to the signal $\psi$ nor relative to one another.   
In practice, however, we want to mimic the response of a \emph{template manifold}, including all physical signals that
include our synthetic nonprecessing binary.    Allowing for similar events at different times, a better diagnostic for
lost signal power due to precession is
\begin{align}
\frac{\rho_{rem,min}^2}{\rho^2} &=\text{min}_t  \frac{\rho_{rem}^2(t)}{\rho^2}
\end{align}
\end{subequations}
where the minimum is over all candidate times $t$ for the merger event specified by $\qmstate{R},\qmstate{L}$.

This ``missing power'' diagnostic behaves qualitatively differently for nonprecessing  and precessing binaries.  
For simplicity, consider a scenario where just the leading-order $(l,m)=(2,\pm 2)$ emission
dominates, so higher harmonics can be omitted to a first approximation, such as  $M=100 M_\odot$ and the initial LIGO noise curve.  
For a nonprecessing binary and for masses dominated by quadrupole emission, the two overlaps $|(R,\hat{\psi}_R)|$ and
$|(L,\hat{\psi}_L)|$ are nearly unity.     Hence, almost no signal power  is lost.
By contrast, precession introduces distinctive phase modulations that cannot be produced by nonprecessing sources
\cite{gwastro-mergers-nr-Alignment-BoyleHarald-2011,gw-astro-SpinAlignedLundgren-FragmentA-Theory}.     Since  precession
occurs during merger, we expect loss of signal amplitude in direct proportion to the phase and amplitude modulations
that precession introduced.   In the language of the above diagnostic, we expect that for precessing binaries the two
overlaps  $|(R,\hat{\psi}_R)|$ and $|(L,\hat{\psi}_L)|$ are substantially less than unity.

\begin{figure*}
\includegraphics[width=0.9\textwidth]{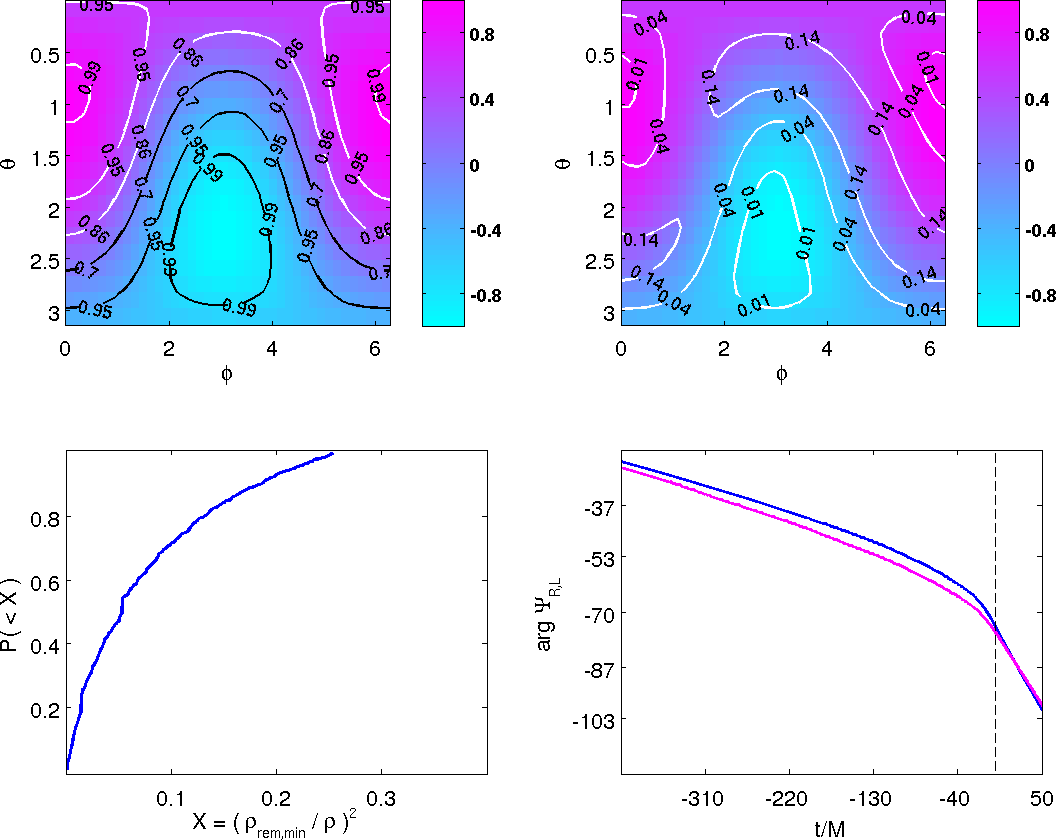}
\caption{\label{fig:PowerLoss:Example}\textbf{Nonprecessing approximation omits signal power}:
Comparison of corotating $(2,\pm 2)$ subspace with the sum of all $l=2$ modes along each line of sight, for  the Sq(4,0.6,90,9) simulation and $M=100
M_\odot$.  
This binary preferentially emits R-handed signal into our detector's sensitive band [$\bar{\rho}_L/\bar{\rho}_R=0.76$
  from Table \ref{tab:Simulations:VPrecession}].
\emph{Top left panel}:%
Contours of normalized matches $|(R,\hat{\psi}_R)|$ (white) and  $|(L,\hat{\psi}_L)|$ (black) for the initial %
 LIGO noise curve at 100Mpc. For comparison, the colors indicate proximity to a single direction ($\hat{n}(\theta,\phi)\cdot \hat{V}^*$ for $\hat{V}^*\equiv \hat{n}(1.01,
2.98)$), the estimated preferred emission direction at $100 M_\odot$. 
The R basis signal is a good match when it
dominates emission and vice-versa.  
\emph{Top right panel}: Contour plot of $\rho_{rem,min}^2/\rho^2$, the fraction of signal power
  lost in a nonprecessing approximation  [Eq. (\ref{eq:rhoLost})].   [This calculation   allows \emph{independent timeshifts in L and R} and
    therefore underestimates the true fraction of power lost.]
In directions nearly perpendicular to the preferred direction  $\hat{V}^*$ defined above, a nonprecessing approximation
fails to capture all available signal information.
\emph{Bottom left panel}: Fraction $P(<)$ of orientations with $\rho_{rem,min}^2/\rho^2$ less than a specific threshold (i.e., which
lose less than a specified fraction of the signal amplitude with a natural nonprecessing approximation).
A significant fraction of all orientations are significantly impacted by precession.
\emph{Bottom right panel}: For the line of sight $(\theta,\phi)=(0.94,0.94)$, the phase of $\WeylScalar_R$ (magenta) and
$\WeylScalar_L$ (blue) are shown.  The two differ by a significant, time-varying phase ($\text{arg}(z)$).  While each polarization 
still resembles a nonprecessing signal, no single nonprecessing signal model can fit both R and L polarizations
simultaneously.  For this reason, along this line of sight a large fraction of the signal power is lost when fitting
with a nonprecessing approximation (top right panel).
}
\end{figure*}

As an example,  Figure \ref{fig:PowerLoss:Example} shows contour plots of these two overlaps (top
panel) and of the total signal amplitude lost (bottom two panels) for a fiducial precessing binary.   In this figure, we
adopt a  \emph{low} reference mass ($M=100 M_\odot$) to ensure both that  $l=2$ emission dominates, and that it occurs along a
relatively stationary  instantaneous  direction; see, e.g., Figure 2 in \cite{gwastro-mergers-nr-Alignment-ROS-IsJEnough}. 
Even in this limit, the bottom left panel of Figure \ref{fig:PowerLoss:Example} shows that most orientations are poorly fit
with an optimal nonprecessing approximation generated from the corotating frame.   
Further, these poorly-fit orientations lie nearly perpendicular to some instantaneous $\hat{V}$ direction.
Finally, in each of these poorly-fit orientations, the \emph{relative phase of L and R} changes significantly.  As seen
in the bottom right panel, while the phase of $\WeylScalar_{R,L}$ \emph{individually} resemble a nonprecessing waveform,
\emph{together} their relative phase evolves in a way we expect that no single nonprecessing system could reproduce.  
Similar behavior occurs for all the precessing binaries we have explored.  
Though we adopt a particularly well-chosen reference signal, this calculation only concretely demonstrates a
this particular precessing binary has many lines of sight where its signal cannot be fit by this particular pair of
nonprecessing basis signals.  
Real nonprecessing searches compare each individual signal with a \emph{template bank} of many candidate nonprecessing
signals.  
 We will address this ``fitting factor'' problem in a subsequent publication that compares our precessing signals to
a family of nonprecessing sources. 
However, motivated by other studies of precession-induced modulations
\cite{gw-astro-SpinAlignedLundgren-FragmentA-Theory}, we anticipate that \emph{any} nonprecessing signal cannot mimic
sufficiently strong geometric precession-induced phase and amplitude modulations.

\begin{figure*}
\includegraphics[width=0.9\textwidth]{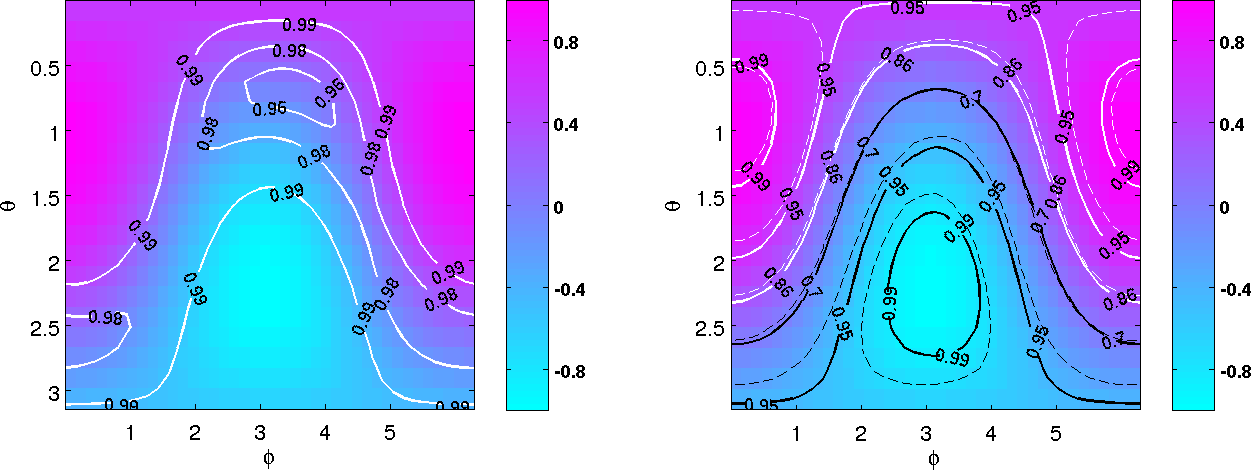}
\caption{\textbf{The line of sight estimate, Eq. \ref{eq:LineofSightEstimate}, is a good approximation for the full $\bf{l=2}$ waveform}: Comparison of $\WeylScalar^{est}$ with the sum of all $l=2$ modes along each line of sight, for  the Sq(4,0.6,90,9) simulation and $M=100
M_\odot$. As before, the colors indicate proximity to a single direction ($\hat{n}(\theta,\phi)\cdot \hat{V}^*$ for $\hat{V}^*\equiv \hat{n}(1.01,
2.98)$), the estimated preferred emission direction at $100 M_\odot$. \emph{Left panel}: Contour of the normalized
match, $|(\WeylScalar^{est},\WeylScalar)|$, for the initial %
 LIGO noise curve at 100Mpc. \emph{Right panel}: Contours of normalized matches
 $|(R,\hat{\psi}^{est})|=|(R,\hat{\psi}^{est}_R)|$ (white) and  $|(L,\hat{\psi}^{est})|=|(L,\hat{\psi}^{est}_L)|$
 (black) for the initial %
LIGO noise curve at 100Mpc. For comparison, the thin dashed lines are the same as those in Fig. \ref{fig:PowerLoss:Example}'s top left panel.
}
\end{figure*}

\subsection{Estimating the mismatch}

Motivated by studies of BH-NS binaries \cite{gw-astro-SpinAlignedLundgren-FragmentA-Theory},  our calculations suggest a
familiar conclusion: when viewed near its  instantaneous orbital plane, a precessing binary can produce complicated
emission, delicately balancing left and right handed power.  No nonprecessing signal model can reproduce it.  
Of course, our signal model also neglects higher harmonics.   Some of the mismatch seen in Figure
\ref{fig:PowerLoss:Example} reflects our inability to recover these other modes.  For the mass region considered,
however, the modes not included produce only a few percent of the (corotating-frame) signal power; see Table 1 in
\cite{gwastro-mergers-nr-Alignment-ROS-IsJEnough}.    The lack of higher harmonics cannot explain the extremely low
matches seen here.

Studies of BH-NS binaries  suggest precession introduces coherent phase and amplitude modulations, superimposed on top of 
secular (corotating-frame) evolution \cite{gw-astro-SpinAlignedLundgren-FragmentA-Theory}.   A sufficiently generic
nonprecessing model can fit any \emph{secular} phase, but not the modulations.  
To show this, we will  rewrite the line-of-sight left- and right-handed signal $\WeylScalar_{R,L}(t)$ as a
modulation factor times a corotating-frame signal.  
Doing so will let us calculate the overlap between corotating-frame modes and any line of sight as a simple integral,
whose  integrand has
two factors: (a) \emph{secular} terms, reflecting the corotating modes' slow change in amplitude with time; and (b)
a \emph{geometrical, precession term} ${\cal X}$, reflecting the change in amplitude and phase along the line of sight.
These expressions prove that the signal power ``lost'' in comparing a nonprecessing to precessing model can
be uniquely associated to the \emph{modulations} that precession introduces in the signal model -- modulations that no
nonprecessing signal can reproduce.
For brevity and for practical reasons -- almost all search strategies use the  $(2,\pm 2)$ modes alone -- we  emphasize a sufficiently low  reference mass that the signal is dominated by  the  $(l,m)=(2,2)$ and $(2,-2)$ modes in its corotating
frame.  
As noted above,  higher order harmonics $l>2$ are still  significant at these masses [Table 1
  in  \cite{gwastro-mergers-nr-Alignment-ROS-IsJEnough}].  
Nonetheless, we explicitly neglect contributions from $l>2$, both to demonstrate our analytic control over the problem
and to connect to  nonprecessing searches.

The line-of-sight gravitational wave signal $\WeylScalar(\hat{n},t)$ can be expressed in terms of   corotating-frame
harmonics $\WeylScalar^{corot}_{lm}$ and a rotation operation:
\begin{eqnarray}
r\WeylScalar(\hat{n},t) &=& \sum_{m,m'}D^2_{mm'}(R(t))\WeylScalar_{m'}^{corot}\Y{-2}_{2,m}(\hat{n})
\end{eqnarray}
where $R(t)$ is a suitable rotation (e.g., the forward transformation $\hat{z}\rightarrow \hat{V}(t)$) and $D^l_{mm'}$ is a Wigner D matrix describing the action of SU(2) rotations on
angular eigenstates.
If the   corotating-frame $(2,\pm 2)$ modes dominate, the line-of-sight signal can be approximated by just two terms:
\begin{eqnarray}
\label{eq:LineofSightEstimate}
r\WeylScalar^{est}(\hat{n},t) &\simeq & \WeylScalar_{22}^{corot} {\cal X}_R + \WeylScalar_{2-2}^{corot} {\cal X}_L  \\
{\cal X}_R &\equiv& \sum_m D^2_{m2}(R(t))\Y{-2}_{2,m}(\hat{n})  \\\
{\cal X}_L &\equiv& \sum_m D^2_{m-2}(R(t))\Y{-2}_{2,m}(\hat{n}) 
\end{eqnarray}
The two slowly-varying but not constant  functions ${\cal X}_{R,L}$ depend only on \emph{geometric} factors: our line of sight $\hat{n}$
and the rotation operator $R$ (i.e., on $\hat{V}$).  

The variation of ${\cal X}_{R,L}$ with time is completely responsible for the structure seen in Figure
\ref{fig:PowerLoss:Example}.    In  Sec. \ref{sec:sub:DemonstrateLossOfSNR}, we described a nonprecessing-signal search strategy, where each
line of sight was projected into the span of two
corotating-frame basis signals $\propto \WeylScalar_{2\pm 2}^{corot}$.   If the two modulation factors ${\cal X}_{R,L}$
were \emph{constant}, these projections would be trivial; the search would recover the line-of-sight signal.  
To estimate the fraction of signal power lost due to precession, we further approximate the right- and left-handed parts
of $\WeylScalar^{est}(\hat{n},t)$  [Eq. (\ref{eq:Chiral})]  by 
\begin{eqnarray}
\WeylScalar_{R}^{est} \simeq \WeylScalar_{22}^{corot} {\cal X}_R \\
\WeylScalar_L^{est} \simeq  \WeylScalar_{2-2}^{corot} {\cal X}_L  
\end{eqnarray}
which is an excellent approximation to the extent that ${\cal X}_{R,L}$ are slowly-varying.   
Given this ansatz, the formulae described in Sec. \ref{sec:sub:DemonstrateLossOfSNR} let us calculate the fraction of
signal-to-noise $\rho_{rem}/\rho$ that a nonprecessing search should miss when applied to this particular line of
sight.   
For example, the overlap $(R,\hat{\psi}_R)$ can be approximated by $(R,\hat{\psi}_R^{est})$, which can in turn be
approximated as
\begin{eqnarray}
(R,\hat{\psi}_R^{est}) &\simeq &  \frac{\qmoperatorelement{R}{{\cal X}_R}{R}
 }{
  \qmoperatorelement{R}{|{\cal X}_R|^2}{R}^{1/2}} \\
 &=& \frac{\qmoperatorelement{\hat{\psi}_R}{{\cal X}_R^{-1,*}}{\hat{\psi}_R}
 }{ 
 \qmoperatorelement{\hat{\psi}_R}{|{\cal X}_R|^{-2}}{\hat{\psi}_{R}}^{1/2}} \\
 & =& \frac{\int_0^\infty  df \frac{ {\cal X}_R |\WeylScalar_{22}^{rot}|^2}{(2\pi f)^4 S_h}   }{
   \sqrt{\int_0^\infty  df \frac{ | {\cal X}_R \WeylScalar_{22}^{rot}|^2}{(2\pi f)^4 S_h}}  } 
\end{eqnarray}
where we have simplified the integral limits using chirality.
A similar expression applies to the other chirality:
\begin{eqnarray}
(L,\hat{\psi}_L^{est}) &\simeq & \frac{\qmoperatorelement{L}{{\cal X}_L}{L} 
 }{
  \qmoperatorelement{L}{|{\cal X}_L|^2}{L}^{1/2}} \\
&=&
\frac{\int_{-\infty}^0 df \frac{ {\cal X}_L |\WeylScalar_{2-2}^{rot}|^2}{(2\pi f)^4  S_h} }{
  \sqrt{\int_{-\infty}^0  df \frac{ |{\cal X}_L\WeylScalar_{22}^{rot}|^2}{(2\pi f)^4 S_h} } } 
\end{eqnarray}
Combined these two approximations  allow us to approximate Fig. \ref{fig:PowerLoss:Example}.  
These approximations suggest that mismatch results from fluctuations in ${\cal X}$ around their median values.   To be
concrete, the total SNR lost due to applying a nonprecessing approximation can be expanded as 
\begin{align}
\rho_{rem}^2 &= \qmoperatorelement{\psi_4}{ [1  - \qmstate{R}\qmstateKet{R} -\qmstate{L}\qmstateKet{L}]}{\psi_4}
\nonumber \\
\label{eq:SNRLost:General}
&= \rho_R^2\left[1- \frac{|\qmoperatorelement{R}{{\cal X}_R}{R}|^2}{\qmoperatorelement{R}{|{\cal X}_R|^2}{R}} \right]
  +\rho_L^2\left[1- \frac{|\qmoperatorelement{L}{{\cal X}_L}{L}|^2}{\qmoperatorelement{L}{|{\cal X}_L|^2}{L}} \right]
\end{align}
For a nonprecessing binary with a fixed preferred direction, the geometrical quantities ${\cal X}$ remain constant.  By
contrast, for a precessing binary, ${\cal X}$ fluctuates, with the fraction of lost power being proportional to rms
fluctuations in ${\cal X}$.

Though simple, these expressions are surprisingly powerful.  In particular, they allow us to estimate \emph{the critical
  threshold for precession, above which its effect is observationally accessible, at a given mass and amplitude}.  
As a simple rule of thumb, a physical process or parameter can impact a Bayesian posterior if the match
($P(A,B)=(A,B)/|A||B|$) changes  by
of order $1/\rho^2$ between $P(A,A)=1$ and $P(A,A+\Delta A)$.  For typical first detections, the signal amplitude   is
expected to be small $\rho \simeq 10$; only features or parameters that change the match by $\gtrsim 1\%$ matter.  
Using nonprecessing waveforms $\psi_{2\pm 2}$ and synthetic precession trajectories $\hat{V}$, we can efficiently
explore what lines of sight and what precession trajectories permit an observationally significant amount of
precession.  We will return to this subject  in a subsequent publication.

As in previous studies of precession  \cite{gw-astro-SpinAlignedLundgren-FragmentA-Theory}, our analysis lends itself to
geometric interpretation and approximation.  
For example, early in the inspiral, the orbital frequency is both well-defined and shorter than all other scales.  A
separation of timescales argument suggests that the time-domain multiplication ${\cal X}_R \qmstate{R}$ could be
implemented as a \emph{frequency-domain} multiplication ${\cal X}_{R,L}(t(f))$.  
In this limit, the inner products $(R,\hat{\psi}_R)$ and $(L,\hat{\psi}_L)$ become (frequency) \emph{averages} over the
signal.  
This approach has been used to calculate fitting factors between nonprecessing searches and precessing black
hole/neutron star binaries \cite{gw-astro-SpinAlignedLundgren-FragmentA-Theory}.  
Further investigation is needed to determine whether a separation-of-timescales approximation remains useful during the
merger and ringdown phase.

\section{Conclusions}
\label{sec:Conclude}

Our paper makes two claims: first, that significant ``orientation'' or ``polarization'' changes occur after the merger of two binary black
holes; and, second, that these changes could be accessible even to present-day detectors, if a source could be detected.

First and foremost, in this paper we report on and provide a simple phenomenological interpretation of  rapid changes in the gravitational
wave emission of merging, precessing black  hole binaries   at and beyond  merger.    
Phenomenologically, the features we report on correspond to coherent, multimodal oscillations in the merger and ringdown
signal that resemble ``precession'' of the remnant black hole.  
We confirm their existence through two time-domain methods: first with ``polarization'', the ratio of left- to
right-handed emission; and second using a preferred direction that traces the direction
of strongest emission.    We provide a simple rule to explain the amplitude and phase of these oscillations in terms of coherent, long-lived
``precession'' of the black hole binary before and the merger remnant after merger. 
Our interpretation agrees with a recently-developed geometric interpretation for high-frequency Kerr quasinormal modes
\cite{gwastro-YangZimmermanEtc-QuasinormalSpectrumAndPrecession2012}.   This correspondence merits further investigation.

Second, using data-analysis-motivated diagnostics, we demonstrate each line of sight carries information about its
orientation relative to the preferred orientation described above.  For the simulations we have performed, we are able
to explicitly show that a nonprecessing search will perform poorly,  in extreme but not uncommon cases losing $\simeq
20\%$ of  signal power even at low mass ($M\simeq 100 M_\odot$).   For sufficiently loud signals, we furthermore suggest that the polarization can be measured
\emph{nonparametrically}.  By analogy  with BH-NS binaries, we expect that the most rapid phase evolution characterizes the binary component
masses and spins, while secular modulations encode the  orientation and spins.  
To summarize, the short signal from merging waveforms carries a surprising amount of information about merger dynamics.
Our analysis suggests that  this information can be phenomenologically extracted, even without a complete and
physically-parameterized model for the waveform.

Our study has significant implications for present data analysis and parameter estimation strategies.  
First and foremost, our proof-of-concept study shows how data analysis strategies can extract easily-understood physics
from complicated, short, multimodal  gravitational wave signals.
Second, our analysis
suggests that nonprecessing models (phenomenological; ringdown; et cetera)  for high-mass sources ($M>100M_\odot$)
can omit significant features and introduce uncontrolled biases when applied to ``typical'' merger events, as typical
events can involve significant post-merger oscillations.   For example, our studies suggest that typical moderate-spin
mergers produce a coherent, multimodal ringdown signal, with significant polarization changes along a typical line of
sight.   
As another example, by analogy to studies of inspiralling
BH-NS binaries \cite{gw-astro-SpinAlignedLundgren-FragmentA-Theory},  nonprecessing searches will be
particularly ineffective for certain lines of sight and spin configurations.  
We will address the selection bias of nonspinning searches for precessing binary mergers in a subsequent publication.  

From our preliminary investigations, we believe the surprisingly rich and rapid behavior of merging, precessing binary
black holes may encode nonlinear features of strong-field gravity.   For example, our analysis suggests that rapid
orientation changes occur at merger, encoding a precession frequency.  We anticipate that  the pre- to post-merger transition
coherently seeds a variety of quasinormal modes, with surprisingly long-lived coherence in the time domain.   
We suggest the relationship between the pre- and post-merger amplitudes of these coherent post-merger
oscillations will provide valuable information about the strong-field merger process.   
All of the features described above can be extracted from simulations and observations  using the techniques described
in this work.  
We therefore suspect that strong polarization changes are observationally accessible features of strong-field gravity
during binary black hole merger.

\clearpage
\appendix

\begin{acknowledgements}
The authors have benefitted from conversations with Evan Ochsner, Chad Hanna, Kip Cannon, Andrew Lundgren,  and Pablo
Laguna.  The authors thank Huan Yang and Aaron Zimmerman for pointing out reference
\cite{gwastro-YangZimmermanEtc-QuasinormalSpectrumAndPrecession2012} and the anonymous referee for a close reading of
the manuscript.    
DS is supported by NSF  awards PHY-0925345, PHY- 0941417, PHY-0903973 and TG-PHY060013N.
ROS is supported by NSF award PHY-0970074, the Bradley Program Fellowship, and the UWM Research
Growth Initiative.
\end{acknowledgements}

\bibliography{paperexport}
\end{document}

%% file: tab-mma-SimulationSet.tex
%
\begin{tabular}{l|rlllllll|l|llllrrr}
Key & $r_{start}/M$ & $q$ & $S_{1,x}/M^2$ & $S_{1,y}/M^2$ & $S_{1,z}/M^2$ & $S_{2,x}/M^2$ & $S_{2,y}/M^2$ & $S_{2,z}/M^2$ &
  $\bar{\rho}_L/\bar{\rho}_R$ &  $\Omega_{VJ}M$ &  $\theta_{VJ,final}$ & $T/M$ & $T_{wave}/M$ & $M/h$\\ 
S(1,0.2,0) & 6.2 & 1 & 0 & 0 & 0.05 & -0.05 & 0 & 0                 & 0.95 & -  &   0.029  & 392 & 250 & 77 \\
S(1,0.2,45) & 6.2 & 1 & 0.0354 & 0. & 0.0354 & -0.05 & 0. & 0.   &0.96  & unk   & 0.006& 380.7 & 246.6 & 77 \\
S(1,0.2,90) & { 6.2} & 1 & 0.05 & 0. & 0. & -0.05 & 0. & 0.      & 1.04   & unk   & - & 389.2 & 217.4  & 77 \\
S(1,0.2,135) &  6.2 & 1 & 0.0354 & 0. & -0.0354 & -0.05 & 0. & 0. & 1.11&  0.55  & 0.01 & 394.9 & 206.9 &  77 \\
S(1,0.2,180) & 6.2 & 1 & 0. & 0. & -0.05 & -0.05 & 0. & 0.            & 1 &  0.07 & 0.036 & 394.9 & 210.8 &  77 \\
S(1,0.2,225) & 6.2 & 1 & -0.0354 & 0. & -0.0354 & -0.05 & 0. & 0.  &1.03 & 0.07 & 0.055& 389.2 & 207.9 &  77 \\
S(1,0.2,270) & 6.2 & 1 & -0.05 & 0. & 0. & -0.05 & 0. & 0.            &1     & 0.08 &0.079& 383.2 & 221.8 &  77 \\
S(1,0.2,315) & 6.2 & 1 & -0.0354 & 0. & 0.0354 & -0.05 & 0. & 0.  &0.98 & 0.08 &0.062 & 384. & 247.6 &  77 \\
S(1,0.4,0) & 6.2 & 1 & 0. & 0. & 0.1 & -0.1 & 0. & 0.                & 0.86    & 0.09 & 0.03 &   398.9 & 277.3 &  77 \\
S(1,0.4,45) & 6.2 & 1 & 0.0707 & 0. & 0.0707 & -0.1 & 0. & 0.   & 0.8       &  -  & & 398.9 & 264.5 &  77 \\
S(1,0.4,90) & { 6.2} & 1 & 0.1 & 0. & 0. & -0.1 & 0. & 0.        &1.09   &  &  -  & 398.9 & 227.1 &  77 \\
S(1,0.4,135) & 6.2 & 1 & 0.0707 & 0. & -0.0707 & -0.1 & 0. & 0.     &1.29& 0.055 & 0.017 & 398.9 & 209.8 &  77 \\
S(1,0.4,180) & 6.2 & 1 & 0. & 0. & -0.1 & -0.1 & 0. & 0.               &1.15& 0.07   & 0.065 & 398.9   & 208.7 &  77 \\
S(1,0.4,225) & 6.2 & 1 & -0.0707 & 0. & -0.0707 & -0.1 & 0. & 0.  &1.06 & 0.07  & 0.11 & 398.9 & 209.6 &  77\\
S(1,0.4,270) & 6.2 & 1 & -0.1 & 0. & 0. & -0.1 & 0. & 0.               &1&   0.08   & 0.16 & 398.9 & 240.8 &  77 \\
S(1,0.4,315) & 6.2 & 1 & -0.0707 & 0. & 0.0707 & -0.1 & 0. & 0.   &0.94 &0.09 & 0.08 &  398.9 & 274.2 &  77 \\
S(1,0.6,0) & 6.2 & 1 & 0. & 0. & 0.15 & -0.15 & 0. & 0.               & 1.07 & 0.1 & 0.068 &  398.9 & 289.7 &  77 \\
S(1,0.6,15) & 6.2 & 1 & 0.0388 & 0. & 0.1449 & -0.15 & 0. & 0.   & 0.95 & 0.1 & 0.05 & 498.9 & 288.7 &  77 \\
S(1,0.6,30) & 6.2 & 1 & 0.075 & 0. & 0.1299 & -0.15 & 0. & 0.     & 0.80 & 0.1 & 0.10&  498.9 & 276.7 &  77 \\
S(1,0.6,45) & 6.2 & 1 & 0.1061 & 0. & 0.1061 & -0.15 & 0. & 0.   &0.69 & 0.1 & 0.047 & 498.9 & 270.5 &  77 \\
S(1,0.6,60) & 6.2 & 1 & 0.1299 & 0. & 0.075 & -0.15 & 0. & 0.    & 0.69 &  0.003  &  0.023& 498.9 & 263.3 &  77 \\
S(1,0.6,75) & 6.2 & 1 & 0.1449 & 0. & 0.0388 & -0.15 & 0. & 0.   &0.84 &  0.08 & 0.012 & 498.9 & 248.2 &  77 \\
S(1,0.6,90) & { 6.2} & 1 & 0.15 & 0. & 0. & -0.15 & 0. & 0.     & 1.15&  -  & - & 398.9 & 229.8 &  77 \\
S(1,0.6,105) & 6.2 & 1 & 0.1449 & 0. & -0.0388 & -0.15 & 0. & 0.  & 1.44 & 0.07   & 0.002 & 498.9 & 200.5 &  77 \\
S(1,0.6,120) & 6.2 & 1 & 0.1299 & 0. & -0.075 & -0.15 & 0. & 0.    & 1.51 & 0.05   & 0.011& 498.9 & 211. &  77 \\
S(1,0.6,135) & 6.2 & 1 & 0.1061 & 0. & -0.1061 & -0.15 & 0. & 0.  & 1.41 & 0.06    & 0.039 & 498.9 & 206.9 &  77 \\
S(1,0.6,150) & 6.2 & 1 & 0.075 & 0. & -0.1299 & -0.15 & 0. & 0.    & 1.27 & 0.06   & 0.049& 498.9 & 202.1 &  77 \\
S(1,0.6,165) & 6.2 & 1 & 0.0388 & 0. & -0.1449 & -0.15 & 0. & 0.  & 1.15 & 0.06    & 0.10 & 498.9 & 198.6 &  77 \\
S(1,0.6,180) & 6.2 & 1 & 0. & 0. & -0.15 & -0.15 & 0. & 0.            & 1.09 & 0.06     & 0.13 & 498.9 & 198.4 &  77 \\
S(1,0.6,195) & 6.2 & 1 & -0.0388 & 0. & -0.1449 & -0.15 & 0. & 0. & 1.07 & 0.06      & 0.13  & 498.9 & 199.8 &  77 \\
S(1,0.6,210) & 6.2 & 1 & -0.075 & 0. & -0.1299 & -0.15 & 0. & 0.   & 1.07 & 0.06-0.07 & 0.14    & 498.9 & 204.6 &  77 \\
S(1,0.6,225) & 6.2 & 1 & -0.1061 & 0. & -0.1061 & -0.15 & 0. & 0  & 1.08 & 0.07      &0.14& 498.9 & 206.5 &  77 \\
S(1,0.6,240) & 6.2 & 1 & -0.1299 & 0. & -0.075 & -0.15 & 0. & 0.   & 1.07 & 0.07      & 0.18 & 498.9 & 204. &  77 \\
S(1,0.6,255) & 6.2 & 1 & -0.1449 & 0. & -0.0388 & -0.15 & 0. & 0. & 1.04 & 0.08      & 0.24 & 495.6 & 233.1 &  77 \\
S(1,0.6,260) & 6.2 & 1 & -0.1477 & 0. & -0.026 & -0.15 & 0. & 0.   & 1.03 & 0.08      & 0.26 & 498.9 & 248. &  77 \\
S(1,0.6,265) & 6.2 & 1 & -0.1494 & 0. & -0.0131 & -0.15 & 0. & 0.  &1.02 & 0.085     & 0.21 & 498.9 & 251.3 &  77 \\
S(1,0.6,270) & 6.2 & 1 & -0.15 & 0. & 0. & -0.15 & 0. & 0.            &1     & 0.09    & 0.15 & 436.7 & 254.4 &  77 \\
S(1,0.6,285) & 6.2 & 1 & -0.1449 & 0. & 0.0388 & -0.15 & 0. & 0.  &0.97 & 0.09    & 0.11& 488.4 & 274.2 &  77 \\
S(1,0.6,300) & 6.2 & 1 & -0.1299 & 0. & 0.075 & -0.15 & 0. & 0.   &0.99 & 0.11     &0.12 & 498.9 & 279.2 &  77 \\
S(1,0.6,315) & 6.2 & 1 & -0.1061 & 0. & 0.1061 & -0.15 & 0. & 0. &1.03  &           & 0.12& 498.9 & 287.7 &  77 \\
S(1,0.6,330) & 6.2 & 1 & -0.075 & 0. & 0.1299 & -0.15 & 0. & 0.   &1.08 & 0.1      & 0.11 & 498.9 & 289.5 &  77 \\
S(1,0.6,345) & 6.2 & 1 & -0.0388 & 0. & 0.1449 & -0.15 & 0. & 0. &1.1   & 0.1      & 0.09 & 498.9 & 290.2 &  77 \\
S(1,0.8,0) &  6.2 & 1 & 0. & 0. & 0.2 & -0.2 & 0. & 0.             &1.5   &  0.12 &0.18 & 498.9 & 306.1  & 77 \\
S(1,0.8,30) &  6.2 & 1 & 0.1 & 0. & 0.1732 & -0.2 & 0. & 0.     &1.24 & 0.11  &0.05 & 498.9 & 292.8  & 77 \\
S(1,0.8,60) &  6.2 & 1 & 0.1732 & 0. & 0.1 & -0.2 & 0. & 0.     &0.59 & 0.097 &0.30 & 498.9 & 269.1  & 77 \\
S(1,0.8,90) &  { 6.2} & 1 & 0.2 & 0. & 0. & -0.2 & 0. & 0.            & 1.09 & - & -& 498.9 & 232.7  & 77 \\  
S(1,0.8,120) &  6.2 & 1 & 0.1732 & 0. & -0.1 & -0.2 & 0. & 0.      &1.70& 0.062 & 0.02 & 498.9 & 210.6  & 77 \\
S(1,0.8,150) &  6.2 & 1 & 0.1 & 0. & -0.1732 & -0.2 & 0. & 0.     & 1.17& 0.061 & 0.097 & 498.9 & 195.3  & 77 \\
S(1,0.8,180) &  6.2 & 1 & 0. & 0. & -0.2 & -0.2 & 0. & 0.          &0.96  & 0.061 & 0.16 & 498.3 & 191.6  & 77 \\
S(1,0.8,210) &  6.2 & 1 & -0.1 & 0. & -0.1732 & -0.2 & 0. & 0.      & 1.04 & 0.069 & 0.19& 498.9 & 202.1  & 77 \\
S(1,0.8,240) &  6.2 & 1 & -0.1732 & 0. & -0.1 & -0.2 & 0. & 0.      & 1.14& 0.081  & 0.25 & 498.9 & 230.4  & 77 \\
S(1,0.8,255) &  6.2 & 1 & -0.1932 & 0. & -0.0518 & -0.2 & 0. & 0. & 1.07 & 0.094 & 0.23 & 498.9 & 255.4  & 77 \\
S(1,0.8,270) &  6.2 & 1 & -0.2 & 0. & 0. & -0.2 & 0. & 0.              & 1 & 0.098    & 0.12 & 498.9 & 277.1  & 77 \\
S(1,0.8,300) &  6.2 & 1 & -0.1732 & 0. & 0.1 & -0.2 & 0. & 0.       & 1.15 & 0.12 &0.17 & 498.9 & 292.  & 77 \\
S(1,0.8,330) &  6.2 & 1 & -0.1 & 0. & 0.1732 & -0.2 & 0. & 0.       & 1.34& 0.12  &0.15 & 498.9 & 306.9 & 77 \\ 
Sq(2,0.6, 0,6.2) &   6.2 & 2 & 0. & 0. & 0.2666 & -0.0666 & 0. & 0.       & 1.11  &  0.12 & 0.034 & 599. & 342.5 & 140 \\
Sq(2,0.6,30,6.2) &  6.2 & 2 & 0.1333 & 0. & 0.2309 & -0.0666 & 0. & 0. & 1.29  & 0.11 &  0.124  & 548. & 326.3 & 140  \\
Sq(2,0.6,90,6.2) &  6.2 & 2 & 0.2666 & 0. & 0. & -0.0666 & 0. & 0.        & 0.96  & 0.07 & 0.12 & 411. & 255.1 & 140  \\
Sq(2,0.6,150,6.2) & 6.2 & 2 & 0.1333 & 0. & -0.2309 & -0.0666 & 0. & 0. &0.99   & 0.04 & 0.093 & 415.2 & 195.4 & 140 \\
Sq(2,0.6,180,6.2) & 6.2 & 2 & 0. & 0. & -0.2666 & -0.0666 & 0. & 0.        & 0.86 &  0.37  & 0.087  & 396.6 & 187.8 & 140 \\
Sq(2,0.6,270,6.2) & 6.2 & 2 & -0.2666 & 0. & 0. & -0.0666 & 0. & 0.        & 1.31  & 0.08  & -    & 497.9 & 267.5 & 140 \\
Sq(4,0.6, 0,6.2) &  6.2 & 4 & 0. & 0. & 0.384 & -0.024 & 0. & 0.             &0.99  & 0.10 & 0.035 & 583.2 & 434.5 & 140 \\
Sq(4,0.6,30,6.2) &  6.2 & 4 & 0.192 & 0. & 0.3325 & -0.024 & 0. & 0.      &1.13  & 0.095& 0.11 & 599. & 427.1 & 140 \\
Sq(4,0.6,90,6.2) & 6.2 & 4 & 0.384 & 0. & 0. & -0.024 & 0. & 0.              &0.96  & 0.058& 0.38  & 549.9 & 264.6 & 140 \\
Sq(4,0.6,150,6.2) &  6.2 & 4 & 0.192 & 0. & -0.3325 & -0.024 & 0. & 0.    &1.06   & 0.01 & 0.55 & 408.1 & 165.6 & 140 \\
Sq(4,0.6,180,6.2) &  6.2 & 4 & 0. & 0. & -0.384 & -0.024 & 0. & 0.           & 0.91  &   - & 0.10 & 415.5 & 165.6 & 140 \\
Sq(4,0.6,270,6.2) &  6.2 & 4 & -0.384 & 0. & 0. & -0.024 & 0. & 0.            & 0.77  & 0.07 & 0.27&  549.9 & 280.6 & 140 \\
Sq(4,0.6,0,9) &  9. & 4 & 0. & 0. & 0.384 & -0.024 & 0. & 0.                 & 1.06  & 0.05-0.1 & 0.022& 1599. & 1273.1 & 140 \\
Sq(4,0.6,90,9) &  9. & 4 & 0.384 & 0. & 0. & -0.024 & 0. & 0.                & 0.76 & 0.058 & 0.33&  1194.4 & 867. & 140 \\
Sq(4,0.6,150,9) &  9. & 4 & -0.192 & 0. & -0.3325 & -0.024 & 0. & 0.      &1.20    &  0.01 & 0.50 & 799. & 487.5 & 140 \\
Sq(4,0.6,180,9) &  9. & 4 & 0. & 0. & -0.384 & -0.024 & 0. & 0.              & 1.1   & - &  0.09 & 799. & 426.1 & 140 \\
Sq(4,0.6,210,9) &  9. & 4 & 0.192 & 0. & -0.3325 & -0.024 & 0. & 0.       & 0.77  & 0.014 & 0.51& 797.5 & 464. & 140 \\
Sq(4,0.6,270,9) &  9. & 4 & -0.384 & 0. & 0. & -0.024 & 0. & 0.            & 1.37 & 0.06 & 0.26 & 1138.7 & 883.2 & 140 \\ 
T(1,0.2,45) &  10& 1 & 0. & 0. & 0.05 & 0.0354 & 0. & 0.0354        & 1.05 &   0.087 & 0.029 & 1298.9 & 849. & 77 \\
T(1,0.2,60) & 10& 1 & 0. & 0. & 0.05 & 0.0433 & 0.  & 0.025          & 1.07 &   0.08  & 0.039& 1298.9 & 837. & 77 \\
T(1,0.2,90) & 10& 1 & 0. & 0. & 0.05 & 0.05 & 0. & 0.                   & 1.03 &   0.07-0.1 & 0.033 &1298.9 & 838.2 & 77 \\
T(1,0.4,45) & 10& 1 & 0. & 0. & 0.1 & 0.0707 & 0. & 0.0707          & 0.96 & 0.1 & 0.031 & 1499. & 986.6 & 77 \\
T(1,0.4,60) & 10& 1 & 0. & 0. & 0.1 & 0.0866 & 0. & 0.05             & 0.86 & 0.1 & - & 1499. & 965.8 & 77 \\
T(1,0.4,90) & 10& 1 & 0. & 0. & 0.1 & 0.1 & 0. & 0.                     & 1.07 & 0.1 & 0.04 & 1499. & 908.3 & 77 \\
T(1,0.6,45) & 10& 1 & 0. & 0. & 0.15 & 0.1061 & 0.  & 0.1061       & 0.92 & 0.11 & 0.18& 1499. & 1052.1 & 77 \\
T(1,0.6,60) & 10& 1 & 0. & 0. & 0.15 & 0.1299 & 0.   & 0.075       & 1.2   &  0.12 & - & 1499. & 993.7 & 77 \\
T(1,0.6,90) & 10& 1 & 0. & 0. & 0.15 & 0.15    & 0.   & 0.            & 0.73  &  0.1 & - & 1499. & 915.7 & 77 \\
T(1,0,0) & 10& 1 & 0. & 0. & 0. & 0. & 0. & 0. & 1 &-  &- & 1099. & 885.5 & 77 \\
T(1,0.2,0) &  10& 1 & 0. & 0. & 0.05 & 0. & 0. & 0.05 & 1 & - &- & 1298.9 & 906.4 & 77 \\
T(1,0.4,0) &  11 & 1 & 0. & 0. & 0.1 & 0. & 0. & 0.1 & 1 & - &- & 1398.9 & 997.5  & 90\\
T(1,0.6,0) & 10& 1 & 0. & 0. & 0.15 & 0. & 0. & 0.15 & 1& - &- & 1499. & 1086.  & 77\\
T(1,0.8,0) &  10& 1 & 0. & 0. & 0.2 & 0. & 0. & 0.2 & 1  & - &- & 1699. & 1217.7  & 90\\ 
Tq(1.5,0.4,60) &  10& 1.5 & 0.1247 & 0. & 0.072 & 0. & 0. & 0.064     & 0.95 & 0.085 &0.055 & 1498.9 & 1157.7 & 120 \\
Tq(1.5,0.6,45) & 10& 1.5 & 0.1527 & 0. & 0.1527 & 0. & 0. & 0.096   & 1.17 & 0.11 &0.027& 1598.9 & 1298.8 & 120 \\
Tq(1.5,0.6,60) & 10& 1.5 & 0.187 & 0. & 0.108 & 0. & 0. & 0.096      & 1.2  & 0.10 & 0.055& 1598.9 & 1251.9 & 120 \\
Tq(1.5,0.6,90) & 10& 1.5 & 0.216 & 0. & 0. & 0. & 0. & 0.096           & 1.34 & 0.07 & -& 1598.8 & 1134.1 & 120 \\
Tq(2,0.4,60) & 10& 2 & 0.1539 & 0. & 0.0889 & 0. & 0. & 0.0444   & 0.91 & 0.08 & 0.084 & 1598.9 & 1216.1 & 120 \\
Tq(2,0.6,45) & 10& 2 & 0.1885 & 0. & 0.1885 & 0. & 0. & 0.0666   &0.90 & 0.106 & 0.068& 1698.9 & 1390.9 & 120 \\
Tq(2,0.6,60) & 10& 2 & 0.2309 & 0. & 0.1333 & 0. & 0. & 0.0666   &0.88 & 0.095 & 0.080& 1698.9 & 1328.4 & 120 \\
Tq(2,0.6,90) & 10& 2 & 0.2666 & 0. & 0. & 0. & 0. & 0.0666          &0.72 & 0.075 & 0.11& 1598.9 & 1176.4 & 120 \\
Tq(2.5,0.4,45) & 10& 2.5 & 0.1443 & 0. & 0.1443 & 0. & 0. & 0.0326 & 0.85& 0.06 &0.072& 1698.9 & 1351.6 & 120 \\
Tq(2.5,0.4,60) & 10& 2.5 & 0.1767 & 0. & 0.102 & 0. & 0. & 0.0326   &1.03 & 0.06 &0.12& 1598.9 & 1297.6 & 120 \\
Tq(2.5,0.4,90) & 10& 2.5 & 0.2041 & 0. & 0. & 0. & 0. & 0.0326        & 0.79& 0.04 &0.12& 1498.9 & 1180.4 & 120 \\
Tq(2.5,0.6,45) & 10& 2.5 & 0.2164 & 0. & 0.2165 & 0. & 0. & 0.049   & 0.94 & 0.095 & 0.073 & 1798.9 & 1488.4 & 120\\
Tq(2.5,0.6,60) & 10& 2.5 & 0.2651 & 0. & 0.1531 & 0. & 0. &0.049   &  1.24  & 0.076& 0.11 &1798.9&1409.6 & 120\\
Tq(2.5,0.6,90) &10 & 2.5 & 0.3061 & 0. & 0. & 0. & 0.& 0.049          & 1.14  & 0.066 &0.20   & 1598.9 &1226.4 &120\\
%
Tq(4, 0.0,0) & 11 & 4 &  0 & 0 & 0 &   0 & 0 & 0 & 1 &  - &- & 2499 & 1989 &  200 \\
Tq(4, 0.6, 45) &10 & 4 & 0.2715 &0. & 0.2715 &0. & 0 & 0.024 & 0.88 &  0.078 & 0.14& 2198.9 &1805.7 & 120\\
Tq(4, 0.6, 60) & 10 &4 & 0.3325 & 0. & 0.192 &0. & 0. & 0.024& 0.90 &  0.058  &0.23& 1998.9 & 1698.1& 120\\
Tq(4, 0.6, 90) & 10 &4 & 0.384  & 0.  & 0 & 0.& 0.    & 0.024 & 1.4 & 0.038      & 0.29& 1798.9 & 1420. &120
\end{tabular}

%% file: paper.bbl
\begin{thebibliography}{59}
\expandafter\ifx\csname natexlab\endcsname\relax\def\natexlab#1{#1}\fi
\expandafter\ifx\csname bibnamefont\endcsname\relax
  \def\bibnamefont#1{#1}\fi
\expandafter\ifx\csname bibfnamefont\endcsname\relax
  \def\bibfnamefont#1{#1}\fi
\expandafter\ifx\csname citenamefont\endcsname\relax
  \def\citenamefont#1{#1}\fi
\expandafter\ifx\csname url\endcsname\relax
  \def\url#1{\texttt{#1}}\fi
\expandafter\ifx\csname urlprefix\endcsname\relax\def\urlprefix{URL }\fi
\providecommand{\bibinfo}[2]{#2}
\providecommand{\eprint}[2][]{\url{#2}}

\bibitem[{\citenamefont{{Abbott et al. (The LIGO Scientific
  Collaboration)}}(2009)}]{gw-detectors-LIGO-original-preferred}
\bibinfo{author}{\bibnamefont{{Abbott et al. (The LIGO Scientific
  Collaboration)}}}, \bibinfo{journal}{Reports on Progress in Physics}
  \textbf{\bibinfo{volume}{72}}, \bibinfo{pages}{076901}
  (\bibinfo{year}{2009}),
  \urlprefix\url{http://stacks.iop.org/0034-4885/72/i=7/a=076901}.

\bibitem[{\citenamefont{{Acernese} and {et
  al}}(2012)}]{gw-detectors-Virgo-original-preferred}
\bibinfo{author}{\bibfnamefont{F.}~\bibnamefont{{Acernese}}} \bibnamefont{and}
  \bibinfo{author}{\bibnamefont{{et al}}}, \bibinfo{journal}{Journal of
  Instrumentation} \textbf{\bibinfo{volume}{7}}, \bibinfo{pages}{P03012}
  (\bibinfo{year}{2012}),
  \urlprefix\url{http://stacks.iop.org/1748-0221/7/i=03/a=P03012}.

\bibitem[{\citenamefont{{M. Punturo et al}}(2010)}]{2010CQGra..27s4002P}
\bibinfo{author}{\bibnamefont{{M. Punturo et al}}}, \bibinfo{journal}{Classical
  and Quantum Gravity} \textbf{\bibinfo{volume}{27}}, \bibinfo{pages}{194002}
  (\bibinfo{year}{2010}).

\bibitem[{\citenamefont{Lehner}(2001)}]{Lehner:2001wq}
\bibinfo{author}{\bibfnamefont{L.}~\bibnamefont{Lehner}},
  \bibinfo{journal}{Class.Quant.Grav.} \textbf{\bibinfo{volume}{18}},
  \bibinfo{pages}{R25} (\bibinfo{year}{2001}), \eprint{gr-qc/0106072}.

\bibitem[{\citenamefont{{Centrella} et~al.}(2010)\citenamefont{{Centrella},
  {Baker}, {Kelly}, and {van Meter}}}]{2010RvMP...82.3069C}
\bibinfo{author}{\bibfnamefont{J.}~\bibnamefont{{Centrella}}},
  \bibinfo{author}{\bibfnamefont{J.~G.} \bibnamefont{{Baker}}},
  \bibinfo{author}{\bibfnamefont{B.~J.} \bibnamefont{{Kelly}}},
  \bibnamefont{and} \bibinfo{author}{\bibfnamefont{J.~R.} \bibnamefont{{van
  Meter}}}, \bibinfo{journal}{Reviews of Modern Physics}
  \textbf{\bibinfo{volume}{82}}, \bibinfo{pages}{3069} (\bibinfo{year}{2010}).

\bibitem[{\citenamefont{Herrmann
  et~al.}(2007{\natexlab{a}})\citenamefont{Herrmann, Hinder, Shoemaker, Laguna,
  and Matzner}}]{Herrmann:2007ex}
\bibinfo{author}{\bibfnamefont{F.}~\bibnamefont{Herrmann}},
  \bibinfo{author}{\bibfnamefont{I.}~\bibnamefont{Hinder}},
  \bibinfo{author}{\bibfnamefont{D.~M.} \bibnamefont{Shoemaker}},
  \bibinfo{author}{\bibfnamefont{P.}~\bibnamefont{Laguna}}, \bibnamefont{and}
  \bibinfo{author}{\bibfnamefont{R.~A.} \bibnamefont{Matzner}},
  \bibinfo{journal}{\prd} \textbf{\bibinfo{volume}{76}},
  \bibinfo{pages}{084032} (\bibinfo{year}{2007}{\natexlab{a}}).

\bibitem[{\citenamefont{Herrmann
  et~al.}(2007{\natexlab{b}})\citenamefont{Herrmann, Hinder, Shoemaker, Laguna,
  and Matzner}}]{Herrmann:2007ac}
\bibinfo{author}{\bibfnamefont{F.}~\bibnamefont{Herrmann}},
  \bibinfo{author}{\bibfnamefont{I.}~\bibnamefont{Hinder}},
  \bibinfo{author}{\bibfnamefont{D.}~\bibnamefont{Shoemaker}},
  \bibinfo{author}{\bibfnamefont{P.}~\bibnamefont{Laguna}}, \bibnamefont{and}
  \bibinfo{author}{\bibfnamefont{R.~A.} \bibnamefont{Matzner}},
  \bibinfo{journal}{Astrophys. J.} \textbf{\bibinfo{volume}{661}},
  \bibinfo{pages}{430} (\bibinfo{year}{2007}{\natexlab{b}}).

\bibitem[{\citenamefont{Healy et~al.}(2009{\natexlab{a}})}]{Healy:2008js}
\bibinfo{author}{\bibfnamefont{J.}~\bibnamefont{Healy}} \bibnamefont{et~al.},
  \bibinfo{journal}{Phys. Rev. Lett.} \textbf{\bibinfo{volume}{102}},
  \bibinfo{pages}{041101} (\bibinfo{year}{2009}{\natexlab{a}}).

\bibitem[{\citenamefont{{Buonanno} et~al.}(2007)\citenamefont{{Buonanno},
  {Cook}, and {Pretorius}}}]{gr-merger-fitting-Alessandra-2006}
\bibinfo{author}{\bibfnamefont{A.}~\bibnamefont{{Buonanno}}},
  \bibinfo{author}{\bibfnamefont{G.~B.} \bibnamefont{{Cook}}},
  \bibnamefont{and}
  \bibinfo{author}{\bibfnamefont{F.}~\bibnamefont{{Pretorius}}},
  \bibinfo{journal}{\prd} \textbf{\bibinfo{volume}{75}},
  \bibinfo{pages}{124018} (\bibinfo{year}{2007}),
  \urlprefix\url{http://xxx.lanl.gov/abs/gr-qc/0610122}.

\bibitem[{\citenamefont{{Abadie et al.\ (LIGO Scientific Collaboration , Virgo
  Collaboration)}}(2011)}]{LIGO-Inspiral-S6-Highmass}
\bibinfo{author}{\bibfnamefont{J.}~\bibnamefont{{Abadie et al.\ (LIGO
  Scientific Collaboration , Virgo Collaboration)}}}, \bibinfo{journal}{\prd}
  \textbf{\bibinfo{volume}{83}}, \bibinfo{pages}{122005}
  (\bibinfo{year}{2011}).

\bibitem[{\citenamefont{{Owen} et~al.}(2011)\citenamefont{{Owen}, {Brink},
  {Chen}, {Kaplan}, {Lovelace}, {Matthews}, {Nichols}, {Scheel}, {Zhang},
  {Zimmerman} et~al.}}]{2011PhRvL.106o1101O}
\bibinfo{author}{\bibfnamefont{R.}~\bibnamefont{{Owen}}},
  \bibinfo{author}{\bibfnamefont{J.}~\bibnamefont{{Brink}}},
  \bibinfo{author}{\bibfnamefont{Y.}~\bibnamefont{{Chen}}},
  \bibinfo{author}{\bibfnamefont{J.~D.} \bibnamefont{{Kaplan}}},
  \bibinfo{author}{\bibfnamefont{G.}~\bibnamefont{{Lovelace}}},
  \bibinfo{author}{\bibfnamefont{K.~D.} \bibnamefont{{Matthews}}},
  \bibinfo{author}{\bibfnamefont{D.~A.} \bibnamefont{{Nichols}}},
  \bibinfo{author}{\bibfnamefont{M.~A.} \bibnamefont{{Scheel}}},
  \bibinfo{author}{\bibfnamefont{F.}~\bibnamefont{{Zhang}}},
  \bibinfo{author}{\bibfnamefont{A.}~\bibnamefont{{Zimmerman}}},
  \bibnamefont{et~al.}, \bibinfo{journal}{Physical Review Letters}
  \textbf{\bibinfo{volume}{106}}, \bibinfo{eid}{151101} (\bibinfo{year}{2011}).

\bibitem[{\citenamefont{{Nichols} et~al.}(2011)\citenamefont{{Nichols}, {Owen},
  {Zhang}, {Zimmerman}, {Brink}, {Chen}, {Kaplan}, {Lovelace}, {Matthews},
  {Scheel} et~al.}}]{2011PhRvD..84l4014N}
\bibinfo{author}{\bibfnamefont{D.~A.} \bibnamefont{{Nichols}}},
  \bibinfo{author}{\bibfnamefont{R.}~\bibnamefont{{Owen}}},
  \bibinfo{author}{\bibfnamefont{F.}~\bibnamefont{{Zhang}}},
  \bibinfo{author}{\bibfnamefont{A.}~\bibnamefont{{Zimmerman}}},
  \bibinfo{author}{\bibfnamefont{J.}~\bibnamefont{{Brink}}},
  \bibinfo{author}{\bibfnamefont{Y.}~\bibnamefont{{Chen}}},
  \bibinfo{author}{\bibfnamefont{J.~D.} \bibnamefont{{Kaplan}}},
  \bibinfo{author}{\bibfnamefont{G.}~\bibnamefont{{Lovelace}}},
  \bibinfo{author}{\bibfnamefont{K.~D.} \bibnamefont{{Matthews}}},
  \bibinfo{author}{\bibfnamefont{M.~A.} \bibnamefont{{Scheel}}},
  \bibnamefont{et~al.}, \bibinfo{journal}{\prd} \textbf{\bibinfo{volume}{84}},
  \bibinfo{eid}{124014} (\bibinfo{year}{2011}).

\bibitem[{\citenamefont{{Christensen} and {Meyer}}(1998)}]{1998PhRvD..58h2001C}
\bibinfo{author}{\bibfnamefont{N.}~\bibnamefont{{Christensen}}}
  \bibnamefont{and} \bibinfo{author}{\bibfnamefont{R.}~\bibnamefont{{Meyer}}},
  \bibinfo{journal}{\prd} \textbf{\bibinfo{volume}{58}},
  \bibinfo{pages}{082001} (\bibinfo{year}{1998}).

\bibitem[{\citenamefont{{Veitch} and {Vecchio}}(2008)}]{2008PhRvD..78b2001V}
\bibinfo{author}{\bibfnamefont{J.}~\bibnamefont{{Veitch}}} \bibnamefont{and}
  \bibinfo{author}{\bibfnamefont{A.}~\bibnamefont{{Vecchio}}},
  \bibinfo{journal}{\prd} \textbf{\bibinfo{volume}{78}},
  \bibinfo{pages}{022001} (\bibinfo{year}{2008}).

\bibitem[{\citenamefont{{R{\"o}ver}
  et~al.}(2007{\natexlab{a}})\citenamefont{{R{\"o}ver}, {Meyer}, {Guidi},
  {Vicer{\'e}}, and {Christensen}}}]{2007CQGra..24..607R}
\bibinfo{author}{\bibfnamefont{C.}~\bibnamefont{{R{\"o}ver}}},
  \bibinfo{author}{\bibfnamefont{R.}~\bibnamefont{{Meyer}}},
  \bibinfo{author}{\bibfnamefont{G.~M.} \bibnamefont{{Guidi}}},
  \bibinfo{author}{\bibfnamefont{A.}~\bibnamefont{{Vicer{\'e}}}},
  \bibnamefont{and}
  \bibinfo{author}{\bibfnamefont{N.}~\bibnamefont{{Christensen}}},
  \bibinfo{journal}{Classical and Quantum Gravity}
  \textbf{\bibinfo{volume}{24}}, \bibinfo{pages}{607}
  (\bibinfo{year}{2007}{\natexlab{a}}).

\bibitem[{\citenamefont{{R{\"o}ver}
  et~al.}(2007{\natexlab{b}})\citenamefont{{R{\"o}ver}, {Meyer}, and
  {Christensen}}}]{2007PhRvD..75f2004R}
\bibinfo{author}{\bibfnamefont{C.}~\bibnamefont{{R{\"o}ver}}},
  \bibinfo{author}{\bibfnamefont{R.}~\bibnamefont{{Meyer}}}, \bibnamefont{and}
  \bibinfo{author}{\bibfnamefont{N.}~\bibnamefont{{Christensen}}},
  \bibinfo{journal}{\prd} \textbf{\bibinfo{volume}{75}},
  \bibinfo{pages}{062004} (\bibinfo{year}{2007}{\natexlab{b}}).

\bibitem[{\citenamefont{{van der Sluys}
  et~al.}(2008{\natexlab{a}})\citenamefont{{van der Sluys}, {R{\"o}ver},
  {Stroeer}, {Raymond}, {Mandel}, {Christensen}, {Kalogera}, {Meyer}, and
  {Vecchio}}}]{2008ApJ...688L..61V}
\bibinfo{author}{\bibfnamefont{M.~V.} \bibnamefont{{van der Sluys}}},
  \bibinfo{author}{\bibfnamefont{C.}~\bibnamefont{{R{\"o}ver}}},
  \bibinfo{author}{\bibfnamefont{A.}~\bibnamefont{{Stroeer}}},
  \bibinfo{author}{\bibfnamefont{V.}~\bibnamefont{{Raymond}}},
  \bibinfo{author}{\bibfnamefont{I.}~\bibnamefont{{Mandel}}},
  \bibinfo{author}{\bibfnamefont{N.}~\bibnamefont{{Christensen}}},
  \bibinfo{author}{\bibfnamefont{V.}~\bibnamefont{{Kalogera}}},
  \bibinfo{author}{\bibfnamefont{R.}~\bibnamefont{{Meyer}}}, \bibnamefont{and}
  \bibinfo{author}{\bibfnamefont{A.}~\bibnamefont{{Vecchio}}},
  \bibinfo{journal}{\apjl} \textbf{\bibinfo{volume}{688}}, \bibinfo{pages}{L61}
  (\bibinfo{year}{2008}{\natexlab{a}}).

\bibitem[{\citenamefont{{van der Sluys}
  et~al.}(2008{\natexlab{b}})\citenamefont{{van der Sluys}, {Raymond},
  {Mandel}, {R{\"o}ver}, {Christensen}, {Kalogera}, {Meyer}, and
  {Vecchio}}}]{2008CQGra..25r4011V}
\bibinfo{author}{\bibfnamefont{M.}~\bibnamefont{{van der Sluys}}},
  \bibinfo{author}{\bibfnamefont{V.}~\bibnamefont{{Raymond}}},
  \bibinfo{author}{\bibfnamefont{I.}~\bibnamefont{{Mandel}}},
  \bibinfo{author}{\bibfnamefont{C.}~\bibnamefont{{R{\"o}ver}}},
  \bibinfo{author}{\bibfnamefont{N.}~\bibnamefont{{Christensen}}},
  \bibinfo{author}{\bibfnamefont{V.}~\bibnamefont{{Kalogera}}},
  \bibinfo{author}{\bibfnamefont{R.}~\bibnamefont{{Meyer}}}, \bibnamefont{and}
  \bibinfo{author}{\bibfnamefont{A.}~\bibnamefont{{Vecchio}}},
  \bibinfo{journal}{Classical and Quantum Gravity}
  \textbf{\bibinfo{volume}{25}}, \bibinfo{pages}{184011}
  (\bibinfo{year}{2008}{\natexlab{b}}).

\bibitem[{\citenamefont{{Cornish} et~al.}(2011)\citenamefont{{Cornish},
  {Sampson}, {Yunes}, and {Pretorius}}}]{2011PhRvD..84f2003C}
\bibinfo{author}{\bibfnamefont{N.}~\bibnamefont{{Cornish}}},
  \bibinfo{author}{\bibfnamefont{L.}~\bibnamefont{{Sampson}}},
  \bibinfo{author}{\bibfnamefont{N.}~\bibnamefont{{Yunes}}}, \bibnamefont{and}
  \bibinfo{author}{\bibfnamefont{F.}~\bibnamefont{{Pretorius}}},
  \bibinfo{journal}{\prd} \textbf{\bibinfo{volume}{84}}, \bibinfo{eid}{062003}
  (\bibinfo{year}{2011}).

\bibitem[{\citenamefont{{Li} et~al.}(2012)\citenamefont{{Li}, {Del Pozzo},
  {Vitale}, {Van Den Broeck}, {Agathos}, {Veitch}, {Grover}, {Sidery},
  {Sturani}, and {Vecchio}}}]{gr-extensions-tests-Europeans2011}
\bibinfo{author}{\bibfnamefont{T.~G.~F.} \bibnamefont{{Li}}},
  \bibinfo{author}{\bibfnamefont{W.}~\bibnamefont{{Del Pozzo}}},
  \bibinfo{author}{\bibfnamefont{S.}~\bibnamefont{{Vitale}}},
  \bibinfo{author}{\bibfnamefont{C.}~\bibnamefont{{Van Den Broeck}}},
  \bibinfo{author}{\bibfnamefont{M.}~\bibnamefont{{Agathos}}},
  \bibinfo{author}{\bibfnamefont{J.}~\bibnamefont{{Veitch}}},
  \bibinfo{author}{\bibfnamefont{K.}~\bibnamefont{{Grover}}},
  \bibinfo{author}{\bibfnamefont{T.}~\bibnamefont{{Sidery}}},
  \bibinfo{author}{\bibfnamefont{R.}~\bibnamefont{{Sturani}}},
  \bibnamefont{and}
  \bibinfo{author}{\bibfnamefont{A.}~\bibnamefont{{Vecchio}}},
  \bibinfo{journal}{\prd} \textbf{\bibinfo{volume}{85}}, \bibinfo{eid}{082003}
  (\bibinfo{year}{2012}), \eprint{1110.0530}.

\bibitem[{\citenamefont{{Del Pozzo} et~al.}(2011)\citenamefont{{Del Pozzo},
  {Veitch}, and {Vecchio}}}]{2011PhRvD..83h2002D}
\bibinfo{author}{\bibfnamefont{W.}~\bibnamefont{{Del Pozzo}}},
  \bibinfo{author}{\bibfnamefont{J.}~\bibnamefont{{Veitch}}}, \bibnamefont{and}
  \bibinfo{author}{\bibfnamefont{A.}~\bibnamefont{{Vecchio}}},
  \bibinfo{journal}{\prd} \textbf{\bibinfo{volume}{83}}, \bibinfo{eid}{082002}
  (\bibinfo{year}{2011}).

\bibitem[{\citenamefont{{Boyle} et~al.}(2008)\citenamefont{{Boyle}, {Kesden},
  and {Nissanke}}}]{gr-nr-io-fitting-Boyle2007}
\bibinfo{author}{\bibfnamefont{L.}~\bibnamefont{{Boyle}}},
  \bibinfo{author}{\bibfnamefont{M.}~\bibnamefont{{Kesden}}}, \bibnamefont{and}
  \bibinfo{author}{\bibfnamefont{S.}~\bibnamefont{{Nissanke}}},
  \bibinfo{journal}{Physical Review Letters} \textbf{\bibinfo{volume}{100}},
  \bibinfo{pages}{151101} (\bibinfo{year}{2008}),
  \urlprefix\url{http://xxx.lanl.gov/abs/arXiv:0709.0299}.

\bibitem[{\citenamefont{{Zlochower} et~al.}(2003)\citenamefont{{Zlochower},
  {G{\'o}mez}, {Husa}, {Lehner}, and {Winicour}}}]{2003PhRvD..68h4014Z}
\bibinfo{author}{\bibfnamefont{Y.}~\bibnamefont{{Zlochower}}},
  \bibinfo{author}{\bibfnamefont{R.}~\bibnamefont{{G{\'o}mez}}},
  \bibinfo{author}{\bibfnamefont{S.}~\bibnamefont{{Husa}}},
  \bibinfo{author}{\bibfnamefont{L.}~\bibnamefont{{Lehner}}}, \bibnamefont{and}
  \bibinfo{author}{\bibfnamefont{J.}~\bibnamefont{{Winicour}}},
  \bibinfo{journal}{\prd} \textbf{\bibinfo{volume}{68}}, \bibinfo{eid}{084014}
  (\bibinfo{year}{2003}), \eprint{arXiv:gr-qc/0306098}.

\bibitem[{\citenamefont{{Nakano} and {Ioka}}(2007)}]{2007PhRvD..76h4007N}
\bibinfo{author}{\bibfnamefont{H.}~\bibnamefont{{Nakano}}} \bibnamefont{and}
  \bibinfo{author}{\bibfnamefont{K.}~\bibnamefont{{Ioka}}},
  \bibinfo{journal}{\prd} \textbf{\bibinfo{volume}{76}}, \bibinfo{eid}{084007}
  (\bibinfo{year}{2007}), \eprint{0708.0450}.

\bibitem[{\citenamefont{{Pazos} et~al.}(2010)\citenamefont{{Pazos}, {Brizuela},
  {Mart{\'{\i}}n-Garc{\'{\i}}a}, and {Tiglio}}}]{2010PhRvD..82j4028P}
\bibinfo{author}{\bibfnamefont{E.}~\bibnamefont{{Pazos}}},
  \bibinfo{author}{\bibfnamefont{D.}~\bibnamefont{{Brizuela}}},
  \bibinfo{author}{\bibfnamefont{J.~M.}
  \bibnamefont{{Mart{\'{\i}}n-Garc{\'{\i}}a}}}, \bibnamefont{and}
  \bibinfo{author}{\bibfnamefont{M.}~\bibnamefont{{Tiglio}}},
  \bibinfo{journal}{\prd} \textbf{\bibinfo{volume}{82}}, \bibinfo{eid}{104028}
  (\bibinfo{year}{2010}), \eprint{1009.4665}.

\bibitem[{\citenamefont{{O'Shaughnessy}
  et~al.}(2011)\citenamefont{{O'Shaughnessy}, {Vaishnav}, {Healy}, {Meeks}, and
  {Shoemaker}}}]{gwastro-mergers-nr-Alignment-ROS-Methods}
\bibinfo{author}{\bibfnamefont{R.}~\bibnamefont{{O'Shaughnessy}}},
  \bibinfo{author}{\bibfnamefont{B.}~\bibnamefont{{Vaishnav}}},
  \bibinfo{author}{\bibfnamefont{J.}~\bibnamefont{{Healy}}},
  \bibinfo{author}{\bibfnamefont{Z.}~\bibnamefont{{Meeks}}}, \bibnamefont{and}
  \bibinfo{author}{\bibfnamefont{D.}~\bibnamefont{{Shoemaker}}},
  \bibinfo{journal}{\prd} \textbf{\bibinfo{volume}{84}},
  \bibinfo{pages}{124002} (\bibinfo{year}{2011}),
  \urlprefix\url{http://link.aps.org/doi/10.1103/PhysRevD.84.124002}.

\bibitem[{\citenamefont{{O'Shaughnessy}
  et~al.}(2012)\citenamefont{{O'Shaughnessy}, {Healy}, {London}, {Meeks}, and
  {Shoemaker}}}]{gwastro-mergers-nr-Alignment-ROS-IsJEnough}
\bibinfo{author}{\bibfnamefont{R.}~\bibnamefont{{O'Shaughnessy}}},
  \bibinfo{author}{\bibfnamefont{J.}~\bibnamefont{{Healy}}},
  \bibinfo{author}{\bibfnamefont{L.}~\bibnamefont{{London}}},
  \bibinfo{author}{\bibfnamefont{Z.}~\bibnamefont{{Meeks}}}, \bibnamefont{and}
  \bibinfo{author}{\bibfnamefont{D.}~\bibnamefont{{Shoemaker}}},
  \bibinfo{journal}{\prd} \textbf{\bibinfo{volume}{85}}, \bibinfo{eid}{084003}
  (\bibinfo{year}{2012}), \urlprefix\url{http://xxx.lanl.gov/abs/1201.2113}.

\bibitem[{\citenamefont{{Yang} et~al.}(2012)\citenamefont{{Yang}, {Nichols},
  {Zhang}, {Zimmerman}, {Zhang}, and
  {Chen}}}]{gwastro-YangZimmermanEtc-QuasinormalSpectrumAndPrecession2012}
\bibinfo{author}{\bibfnamefont{H.}~\bibnamefont{{Yang}}},
  \bibinfo{author}{\bibfnamefont{D.~A.} \bibnamefont{{Nichols}}},
  \bibinfo{author}{\bibfnamefont{F.}~\bibnamefont{{Zhang}}},
  \bibinfo{author}{\bibfnamefont{A.}~\bibnamefont{{Zimmerman}}},
  \bibinfo{author}{\bibfnamefont{Z.}~\bibnamefont{{Zhang}}}, \bibnamefont{and}
  \bibinfo{author}{\bibfnamefont{Y.}~\bibnamefont{{Chen}}},
  \bibinfo{journal}{ArXiv e-prints}  (\bibinfo{year}{2012}),
  \eprint{1207.4253}.

\bibitem[{\citenamefont{Hinder et~al.}(2008)\citenamefont{Hinder, Vaishnav,
  Herrmann, Shoemaker, and Laguna}}]{Hinder:2007qu}
\bibinfo{author}{\bibfnamefont{I.}~\bibnamefont{Hinder}},
  \bibinfo{author}{\bibfnamefont{B.}~\bibnamefont{Vaishnav}},
  \bibinfo{author}{\bibfnamefont{F.}~\bibnamefont{Herrmann}},
  \bibinfo{author}{\bibfnamefont{D.}~\bibnamefont{Shoemaker}},
  \bibnamefont{and} \bibinfo{author}{\bibfnamefont{P.}~\bibnamefont{Laguna}},
  \bibinfo{journal}{\prd} \textbf{\bibinfo{volume}{77}},
  \bibinfo{pages}{081502} (\bibinfo{year}{2008}).

\bibitem[{\citenamefont{{Hinder} et~al.}(2010)\citenamefont{{Hinder},
  {Herrmann}, {Laguna}, and {Shoemaker}}}]{Hinder:2008kv}
\bibinfo{author}{\bibfnamefont{I.}~\bibnamefont{{Hinder}}},
  \bibinfo{author}{\bibfnamefont{F.}~\bibnamefont{{Herrmann}}},
  \bibinfo{author}{\bibfnamefont{P.}~\bibnamefont{{Laguna}}}, \bibnamefont{and}
  \bibinfo{author}{\bibfnamefont{D.}~\bibnamefont{{Shoemaker}}},
  \bibinfo{journal}{\prd} \textbf{\bibinfo{volume}{82}}, \bibinfo{eid}{024033}
  (\bibinfo{year}{2010}), \eprint{0806.1037}.

\bibitem[{\citenamefont{Healy et~al.}(2009{\natexlab{b}})\citenamefont{Healy,
  Levin, and Shoemaker}}]{Healy:2009zm}
\bibinfo{author}{\bibfnamefont{J.}~\bibnamefont{Healy}},
  \bibinfo{author}{\bibfnamefont{J.}~\bibnamefont{Levin}}, \bibnamefont{and}
  \bibinfo{author}{\bibfnamefont{D.}~\bibnamefont{Shoemaker}},
  \bibinfo{journal}{Phys. Rev. Lett.} \textbf{\bibinfo{volume}{103}},
  \bibinfo{pages}{131101} (\bibinfo{year}{2009}{\natexlab{b}}).

\bibitem[{\citenamefont{Healy et~al.}(2010)\citenamefont{Healy, Laguna,
  Matzner, and Shoemaker}}]{Healy:2009ir}
\bibinfo{author}{\bibfnamefont{J.}~\bibnamefont{Healy}},
  \bibinfo{author}{\bibfnamefont{P.}~\bibnamefont{Laguna}},
  \bibinfo{author}{\bibfnamefont{R.~A.} \bibnamefont{Matzner}},
  \bibnamefont{and} \bibinfo{author}{\bibfnamefont{D.~M.}
  \bibnamefont{Shoemaker}}, \bibinfo{journal}{\prd}
  \textbf{\bibinfo{volume}{81}}, \bibinfo{pages}{081501}
  (\bibinfo{year}{2010}).

\bibitem[{\citenamefont{Bode et~al.}(2010)\citenamefont{Bode, Haas, Bogdanovic,
  Laguna, and Shoemaker}}]{Bode:2009mt}
\bibinfo{author}{\bibfnamefont{T.}~\bibnamefont{Bode}},
  \bibinfo{author}{\bibfnamefont{R.}~\bibnamefont{Haas}},
  \bibinfo{author}{\bibfnamefont{T.}~\bibnamefont{Bogdanovic}},
  \bibinfo{author}{\bibfnamefont{P.}~\bibnamefont{Laguna}}, \bibnamefont{and}
  \bibinfo{author}{\bibfnamefont{D.}~\bibnamefont{Shoemaker}},
  \bibinfo{journal}{Astrophys. J.} \textbf{\bibinfo{volume}{715}},
  \bibinfo{pages}{1117} (\bibinfo{year}{2010}).

\bibitem[{\citenamefont{Schnetter et~al.}(2004)\citenamefont{Schnetter, Hawley,
  and Hawke}}]{Schnetter-etal-03b}
\bibinfo{author}{\bibfnamefont{E.}~\bibnamefont{Schnetter}},
  \bibinfo{author}{\bibfnamefont{S.~H.} \bibnamefont{Hawley}},
  \bibnamefont{and} \bibinfo{author}{\bibfnamefont{I.}~\bibnamefont{Hawke}},
  \bibinfo{journal}{Class. Quant. Grav.} \textbf{\bibinfo{volume}{21}},
  \bibinfo{pages}{1465} (\bibinfo{year}{2004}).

\bibitem[{cactus-web()}]{cactus-web}
cactus-web, \bibinfo{note}{cactus Computational Toolkit home page:\\{\tt
  http://www.cactuscode.org}}.

\bibitem[{\citenamefont{Husa et~al.}(2006)\citenamefont{Husa, Hinder, and
  Lechner}}]{Husa:2004ip}
\bibinfo{author}{\bibfnamefont{S.}~\bibnamefont{Husa}},
  \bibinfo{author}{\bibfnamefont{I.}~\bibnamefont{Hinder}}, \bibnamefont{and}
  \bibinfo{author}{\bibfnamefont{C.}~\bibnamefont{Lechner}},
  \bibinfo{journal}{Computer Physics Communications}
  \textbf{\bibinfo{volume}{174}}, \bibinfo{pages}{983} (\bibinfo{year}{2006}).

\bibitem[{\citenamefont{{Wiaux} et~al.}(2007)\citenamefont{{Wiaux}, {Jacques},
  and {Vandergheynst}}}]{2007JCoPh.226.2359W}
\bibinfo{author}{\bibfnamefont{Y.}~\bibnamefont{{Wiaux}}},
  \bibinfo{author}{\bibfnamefont{L.}~\bibnamefont{{Jacques}}},
  \bibnamefont{and}
  \bibinfo{author}{\bibfnamefont{P.}~\bibnamefont{{Vandergheynst}}},
  \bibinfo{journal}{Journal of Computational Physics}
  \textbf{\bibinfo{volume}{226}}, \bibinfo{pages}{2359} (\bibinfo{year}{2007}),
  \eprint{arXiv:astro-ph/0508514}.

\bibitem[{\citenamefont{{Dreyer} et~al.}(2003)\citenamefont{{Dreyer},
  {Krishnan}, {Shoemaker}, and {Schnetter}}}]{2003PhRvD..67b4018D}
\bibinfo{author}{\bibfnamefont{O.}~\bibnamefont{{Dreyer}}},
  \bibinfo{author}{\bibfnamefont{B.}~\bibnamefont{{Krishnan}}},
  \bibinfo{author}{\bibfnamefont{D.}~\bibnamefont{{Shoemaker}}},
  \bibnamefont{and}
  \bibinfo{author}{\bibfnamefont{E.}~\bibnamefont{{Schnetter}}},
  \bibinfo{journal}{\prd} \textbf{\bibinfo{volume}{67}}, \bibinfo{eid}{024018}
  (\bibinfo{year}{2003}), \eprint{arXiv:gr-qc/0206008}.

\bibitem[{\citenamefont{{Ashtekar} and {Krishnan}}(2004)}]{2004LRR.....7...10A}
\bibinfo{author}{\bibfnamefont{A.}~\bibnamefont{{Ashtekar}}} \bibnamefont{and}
  \bibinfo{author}{\bibfnamefont{B.}~\bibnamefont{{Krishnan}}},
  \bibinfo{journal}{Living Reviews in Relativity} \textbf{\bibinfo{volume}{7}},
  \bibinfo{pages}{10} (\bibinfo{year}{2004}), \eprint{arXiv:gr-qc/0407042}.

\bibitem[{\citenamefont{{Schnetter} et~al.}(2006)\citenamefont{{Schnetter},
  {Krishnan}, and {Beyer}}}]{2006PhRvD..74b4028S}
\bibinfo{author}{\bibfnamefont{E.}~\bibnamefont{{Schnetter}}},
  \bibinfo{author}{\bibfnamefont{B.}~\bibnamefont{{Krishnan}}},
  \bibnamefont{and} \bibinfo{author}{\bibfnamefont{F.}~\bibnamefont{{Beyer}}},
  \bibinfo{journal}{\prd} \textbf{\bibinfo{volume}{74}}, \bibinfo{eid}{024028}
  (\bibinfo{year}{2006}), \eprint{arXiv:gr-qc/0604015}.

\bibitem[{\citenamefont{Apostolatos et~al.}(1994)\citenamefont{Apostolatos,
  Cutler, Sussman, and Thorne}}]{ACST}
\bibinfo{author}{\bibfnamefont{T.~A.} \bibnamefont{Apostolatos}},
  \bibinfo{author}{\bibfnamefont{C.}~\bibnamefont{Cutler}},
  \bibinfo{author}{\bibfnamefont{G.~J.} \bibnamefont{Sussman}},
  \bibnamefont{and} \bibinfo{author}{\bibfnamefont{K.~S.}
  \bibnamefont{Thorne}}, \bibinfo{journal}{\prd} \textbf{\bibinfo{volume}{49}},
  \bibinfo{pages}{6274} (\bibinfo{year}{1994}).

\bibitem[{\citenamefont{{Ochsner} and
  {O'Shaughnessy}}(2012)}]{gwastro-mergers-nr-Alignment-ROS-PN}
\bibinfo{author}{\bibfnamefont{E.}~\bibnamefont{{Ochsner}}} \bibnamefont{and}
  \bibinfo{author}{\bibfnamefont{R.}~\bibnamefont{{O'Shaughnessy}}},
  \bibinfo{journal}{in LSC review (DCC P120043); to be submitted to PRD}
  (\bibinfo{year}{2012}).

\bibitem[{\citenamefont{{Schmidt} et~al.}(2011)\citenamefont{{Schmidt},
  {Hannam}, {Husa}, and
  {Ajith}}}]{gwastro-mergers-nr-ComovingFrameExpansionSchmidt2010}
\bibinfo{author}{\bibfnamefont{P.}~\bibnamefont{{Schmidt}}},
  \bibinfo{author}{\bibfnamefont{M.}~\bibnamefont{{Hannam}}},
  \bibinfo{author}{\bibfnamefont{S.}~\bibnamefont{{Husa}}}, \bibnamefont{and}
  \bibinfo{author}{\bibfnamefont{P.}~\bibnamefont{{Ajith}}},
  \bibinfo{journal}{\prd} \textbf{\bibinfo{volume}{84}}, \bibinfo{eid}{024046}
  (\bibinfo{year}{2011}),
  \urlprefix\url{http://xxx.lanl.gov/abs/arXiv:1012.2879}.

\bibitem[{\citenamefont{{Boyle} et~al.}(2011)\citenamefont{{Boyle}, {Owen}, and
  {Pfeiffer}}}]{gwastro-mergers-nr-Alignment-BoyleHarald-2011}
\bibinfo{author}{\bibfnamefont{M.}~\bibnamefont{{Boyle}}},
  \bibinfo{author}{\bibfnamefont{R.}~\bibnamefont{{Owen}}}, \bibnamefont{and}
  \bibinfo{author}{\bibfnamefont{H.~P.} \bibnamefont{{Pfeiffer}}},
  \bibinfo{journal}{\prd} \textbf{\bibinfo{volume}{84}}, \bibinfo{eid}{124011}
  (\bibinfo{year}{2011}).

\bibitem[{\citenamefont{{Brown} et~al.}(2012)\citenamefont{{Brown}, {Lundgren},
  and {O'Shaughnessy}}}]{gw-astro-SpinAlignedLundgren-FragmentA-Theory}
\bibinfo{author}{\bibfnamefont{D.~A.} \bibnamefont{{Brown}}},
  \bibinfo{author}{\bibfnamefont{A.}~\bibnamefont{{Lundgren}}},
  \bibnamefont{and}
  \bibinfo{author}{\bibfnamefont{R.}~\bibnamefont{{O'Shaughnessy}}},
  \bibinfo{journal}{\prd} \textbf{\bibinfo{volume}{86}}, \bibinfo{eid}{064020}
  (\bibinfo{year}{2012}), \eprint{1203.6060},
  \urlprefix\url{http://arxiv.org/abs/1203.6060}.

\bibitem[{\citenamefont{{Br{\"u}gmann}
  et~al.}(2008)\citenamefont{{Br{\"u}gmann}, {Gonz{\'a}lez}, {Hannam}, {Husa},
  and {Sperhake}}}]{2008PhRvD..77l4047B}
\bibinfo{author}{\bibfnamefont{B.}~\bibnamefont{{Br{\"u}gmann}}},
  \bibinfo{author}{\bibfnamefont{J.~A.} \bibnamefont{{Gonz{\'a}lez}}},
  \bibinfo{author}{\bibfnamefont{M.}~\bibnamefont{{Hannam}}},
  \bibinfo{author}{\bibfnamefont{S.}~\bibnamefont{{Husa}}}, \bibnamefont{and}
  \bibinfo{author}{\bibfnamefont{U.}~\bibnamefont{{Sperhake}}},
  \bibinfo{journal}{\prd} \textbf{\bibinfo{volume}{77}}, \bibinfo{eid}{124047}
  (\bibinfo{year}{2008}), \eprint{0707.0135}.

\bibitem[{\citenamefont{{Lousto} et~al.}(2010)\citenamefont{{Lousto},
  {Campanelli}, {Zlochower}, and {Nakano}}}]{2010CQGra..27k4006L}
\bibinfo{author}{\bibfnamefont{C.~O.} \bibnamefont{{Lousto}}},
  \bibinfo{author}{\bibfnamefont{M.}~\bibnamefont{{Campanelli}}},
  \bibinfo{author}{\bibfnamefont{Y.}~\bibnamefont{{Zlochower}}},
  \bibnamefont{and} \bibinfo{author}{\bibfnamefont{H.}~\bibnamefont{{Nakano}}},
  \bibinfo{journal}{Classical and Quantum Gravity}
  \textbf{\bibinfo{volume}{27}}, \bibinfo{pages}{114006}
  (\bibinfo{year}{2010}).

\bibitem[{\citenamefont{{O'Shaughnessy}
  et~al.}(2010)\citenamefont{{O'Shaughnessy}, {Vaishnav}, {Healy}, and
  {Shoemaker}}}]{gwastro-spins-rangefit2010}
\bibinfo{author}{\bibfnamefont{R.}~\bibnamefont{{O'Shaughnessy}}},
  \bibinfo{author}{\bibfnamefont{B.}~\bibnamefont{{Vaishnav}}},
  \bibinfo{author}{\bibfnamefont{J.}~\bibnamefont{{Healy}}}, \bibnamefont{and}
  \bibinfo{author}{\bibfnamefont{D.}~\bibnamefont{{Shoemaker}}},
  \bibinfo{journal}{\prd} \textbf{\bibinfo{volume}{82}},
  \bibinfo{pages}{104006} (\bibinfo{year}{2010}), \eprint{(arXiv:1007.4213)},
  \urlprefix\url{http://xxx.lanl.gov/abs/arXiv:1007.4213}.

\bibitem[{\citenamefont{{Pekowsky} et~al.}(2013)\citenamefont{{Pekowsky},
  Healy, {London}, {O'Shaughnessy}, and
  {Shoemaker}}}]{gwastro-mergers-nr-Alignment-ROS-CorotatingWaveforms}
\bibinfo{author}{\bibfnamefont{L.}~\bibnamefont{{Pekowsky}}},
  \bibinfo{author}{\bibfnamefont{J.}~\bibnamefont{Healy}},
  \bibinfo{author}{\bibfnamefont{L.}~\bibnamefont{{London}}},
  \bibinfo{author}{\bibfnamefont{R.}~\bibnamefont{{O'Shaughnessy}}},
  \bibnamefont{and}
  \bibinfo{author}{\bibfnamefont{D.}~\bibnamefont{{Shoemaker}}},
  \bibinfo{journal}{in preparation}  (\bibinfo{year}{2013}).

\bibitem[{\citenamefont{{Berti} et~al.}(2009)\citenamefont{{Berti}, {Cardoso},
  and {Starinets}}}]{2009CQGra..26p3001B}
\bibinfo{author}{\bibfnamefont{E.}~\bibnamefont{{Berti}}},
  \bibinfo{author}{\bibfnamefont{V.}~\bibnamefont{{Cardoso}}},
  \bibnamefont{and} \bibinfo{author}{\bibfnamefont{A.~O.}
  \bibnamefont{{Starinets}}}, \bibinfo{journal}{Classical and Quantum Gravity}
  \textbf{\bibinfo{volume}{26}}, \bibinfo{pages}{163001}
  (\bibinfo{year}{2009}), \eprint{0905.2975}.

\bibitem[{\citenamefont{{Poisson} and {Will}}(1995)}]{1995PhRvD..52..848P}
\bibinfo{author}{\bibfnamefont{E.}~\bibnamefont{{Poisson}}} \bibnamefont{and}
  \bibinfo{author}{\bibfnamefont{C.~M.} \bibnamefont{{Will}}},
  \bibinfo{journal}{\prd} \textbf{\bibinfo{volume}{52}}, \bibinfo{pages}{848}
  (\bibinfo{year}{1995}), \eprint{arXiv:gr-qc/9502040}.

\bibitem[{\citenamefont{{Vallisneri}}(2008)}]{gw-astro-Vallis-Fisher-2007}
\bibinfo{author}{\bibfnamefont{M.}~\bibnamefont{{Vallisneri}}},
  \bibinfo{journal}{\prd} \textbf{\bibinfo{volume}{77}}, \bibinfo{eid}{042001}
  (\bibinfo{year}{2008}), \eprint{arXiv:gr-qc/0703086}.

\bibitem[{\citenamefont{{Cutler} and {Flanagan}}(1994)}]{CutlerFlanagan:1994}
\bibinfo{author}{\bibfnamefont{C.}~\bibnamefont{{Cutler}}} \bibnamefont{and}
  \bibinfo{author}{\bibfnamefont{E.~E.} \bibnamefont{{Flanagan}}},
  \bibinfo{journal}{\prd} \textbf{\bibinfo{volume}{49}}, \bibinfo{pages}{2658}
  (\bibinfo{year}{1994}), \eprint{gr-qc/9402014}.

\bibitem[{\citenamefont{{Raymond} et~al.}(2009)\citenamefont{{Raymond}, {van
  der Sluys}, {Mandel}, {Kalogera}, {R{\"o}ver}, and
  {Christensen}}}]{2009CQGra..26k4007R}
\bibinfo{author}{\bibfnamefont{V.}~\bibnamefont{{Raymond}}},
  \bibinfo{author}{\bibfnamefont{M.~V.} \bibnamefont{{van der Sluys}}},
  \bibinfo{author}{\bibfnamefont{I.}~\bibnamefont{{Mandel}}},
  \bibinfo{author}{\bibfnamefont{V.}~\bibnamefont{{Kalogera}}},
  \bibinfo{author}{\bibfnamefont{C.}~\bibnamefont{{R{\"o}ver}}},
  \bibnamefont{and}
  \bibinfo{author}{\bibfnamefont{N.}~\bibnamefont{{Christensen}}},
  \bibinfo{journal}{Classical and Quantum Gravity}
  \textbf{\bibinfo{volume}{26}}, \bibinfo{pages}{114007}
  (\bibinfo{year}{2009}), \eprint{0812.4302}.

\bibitem[{\citenamefont{{Veitch} et~al.}(2012)\citenamefont{{Veitch}, {Mandel},
  {Aylott}, {Farr}, {Raymond}, {Rodriguez}, {van der Sluys}, {Kalogera}, and
  {Vecchio}}}]{gwastro-mergers-PE-Aylott-LIGOATest}
\bibinfo{author}{\bibfnamefont{J.}~\bibnamefont{{Veitch}}},
  \bibinfo{author}{\bibfnamefont{I.}~\bibnamefont{{Mandel}}},
  \bibinfo{author}{\bibfnamefont{B.}~\bibnamefont{{Aylott}}},
  \bibinfo{author}{\bibfnamefont{B.}~\bibnamefont{{Farr}}},
  \bibinfo{author}{\bibfnamefont{V.}~\bibnamefont{{Raymond}}},
  \bibinfo{author}{\bibfnamefont{C.}~\bibnamefont{{Rodriguez}}},
  \bibinfo{author}{\bibfnamefont{M.}~\bibnamefont{{van der Sluys}}},
  \bibinfo{author}{\bibfnamefont{V.}~\bibnamefont{{Kalogera}}},
  \bibnamefont{and}
  \bibinfo{author}{\bibfnamefont{A.}~\bibnamefont{{Vecchio}}},
  \bibinfo{journal}{\prd} \textbf{\bibinfo{volume}{85}}, \bibinfo{eid}{104045}
  (\bibinfo{year}{2012}), \eprint{1201.1195}.

\bibitem[{\citenamefont{{Nissanke} et~al.}(2011)\citenamefont{{Nissanke},
  {Sievers}, {Dalal}, and {Holz}}}]{2011ApJ...739...99N}
\bibinfo{author}{\bibfnamefont{S.}~\bibnamefont{{Nissanke}}},
  \bibinfo{author}{\bibfnamefont{J.}~\bibnamefont{{Sievers}}},
  \bibinfo{author}{\bibfnamefont{N.}~\bibnamefont{{Dalal}}}, \bibnamefont{and}
  \bibinfo{author}{\bibfnamefont{D.}~\bibnamefont{{Holz}}},
  \bibinfo{journal}{\apj} \textbf{\bibinfo{volume}{739}}, \bibinfo{eid}{99}
  (\bibinfo{year}{2011}), \eprint{1105.3184}.

\bibitem[{\citenamefont{{Lang} et~al.}(2011)\citenamefont{{Lang}, {Hughes}, and
  {Cornish}}}]{2011PhRvD..84b2002L}
\bibinfo{author}{\bibfnamefont{R.~N.} \bibnamefont{{Lang}}},
  \bibinfo{author}{\bibfnamefont{S.~A.} \bibnamefont{{Hughes}}},
  \bibnamefont{and} \bibinfo{author}{\bibfnamefont{N.~J.}
  \bibnamefont{{Cornish}}}, \bibinfo{journal}{\prd}
  \textbf{\bibinfo{volume}{84}}, \bibinfo{eid}{022002} (\bibinfo{year}{2011}),
  \eprint{1101.3591}.

\bibitem[{\citenamefont{{Klimenko} et~al.}(2005)\citenamefont{{Klimenko},
  {Mohanty}, {Rakhmanov}, and {Mitselmakher}}}]{CWaveburst1}
\bibinfo{author}{\bibfnamefont{S.}~\bibnamefont{{Klimenko}}},
  \bibinfo{author}{\bibfnamefont{S.}~\bibnamefont{{Mohanty}}},
  \bibinfo{author}{\bibfnamefont{M.}~\bibnamefont{{Rakhmanov}}},
  \bibnamefont{and}
  \bibinfo{author}{\bibfnamefont{G.}~\bibnamefont{{Mitselmakher}}},
  \bibinfo{journal}{{Phys. Rev. D}} \textbf{\bibinfo{volume}{72}},
  \bibinfo{pages}{122002} (\bibinfo{year}{2005}),
  \urlprefix\url{http://arxiv.org/abs/gr-qc/0508068}.

\bibitem[{\citenamefont{{Klimenko} et~al.}(2008)\citenamefont{{Klimenko},
  {Yakushin}, {Mercer}, and {Mitselmakher}}}]{2008CQGra..25k4029K}
\bibinfo{author}{\bibfnamefont{S.}~\bibnamefont{{Klimenko}}},
  \bibinfo{author}{\bibfnamefont{I.}~\bibnamefont{{Yakushin}}},
  \bibinfo{author}{\bibfnamefont{A.}~\bibnamefont{{Mercer}}}, \bibnamefont{and}
  \bibinfo{author}{\bibfnamefont{G.}~\bibnamefont{{Mitselmakher}}},
  \bibinfo{journal}{Classical and Quantum Gravity}
  \textbf{\bibinfo{volume}{25}}, \bibinfo{pages}{114029}
  (\bibinfo{year}{2008}).

\end{thebibliography}
